%% file: HIG-23-014_temp.tex
\pdfoutput=1
\documentclass[11pt,twoside,a4paper,cmspaper,final,collab]{cms-tdr}
\def\svnVersion{ce79a6e}\def\svnDate{2025/09/13}\def\cmsCernNoTag{CERN-EP-2025-067}\def\cmsCernDate{\today}\def\cmsMessage{Published in the Journal of High Energy Physics as \href{http://dx.doi.org/10.1007/JHEP09(2025)070}{\doi{10.1007/JHEP09(2025)070}.}}
\begin{document}\cmsNoteHeader{HIG-23-014}

\newcommand{\sqrtstpstev}{\ensuremath{\sqrt{s}=13.6\TeV}\xspace}
\newcommand{\sqrtsttev}{\ensuremath{\sqrt{s}=13\TeV}\xspace} 
\newcommand{\fb}{\ensuremath{\unit{fb}}\xspace}
\newcommand{\Hgg}{\ensuremath{\PH\to\PGg\PGg}\xspace}
\newcommand{\HZZStarFourEll}{\ensuremath{\PH\to \PZ\PZ\to 4\ell}\xspace}
\newcommand{\HWWStarEllPrimed}{\ensuremath{\PH \to \PW\PW \to \ell \ell' \PGn \PGn'}\xspace}
\newcommand{\Htautau}{\ensuremath{\PH\to\PGt\PGt}\xspace}
\newcommand{\Hbb}{\ensuremath{\PH\to\PQb\PQb}\xspace}
\newcommand{\mgg}{\ensuremath{m_{\PGg\PGg}}\xspace}    
\newcommand{\sigFid}{\ensuremath{\sigma_{\text{fid}}}\xspace}
\newcommand{\Zee}{\ensuremath{\PZ \to \Pe\Pe }\xspace}  
\newcommand{\gamOne}{\ensuremath{\PGg_1}\xspace}
\newcommand{\gamTwo}{\ensuremath{\PGg_2}\xspace}
\newcommand{\ggH}{ggH\xspace}
\newcommand{\VBF}{VBF\xspace}
\newcommand{\VH}{VH\xspace}
\newcommand{\ttH}{\ensuremath{\ttbar\PH}\xspace}
\newcommand{\NNLOPS}{\textsc{nnlops}\xspace}
\newcommand{\gamgam}{\ensuremath{\PGg\PGg}\xspace}
\newcommand{\gjet}{\ensuremath{\PGg+\text{jet}}\xspace}
\newcommand{\Zmmg}{\ensuremath{\PZ\to\PGm\PGm\PGg}\xspace}
\newcommand{\mb}{\ensuremath{\unit{mb}}\xspace}
\newcommand{\HoE}{\ensuremath{H/E}\xspace}
\newcommand{\ptGam}{\ensuremath{\pt^{\PGg}}\xspace}
\newcommand{\gamProbee}{\ensuremath{\PGg_{\text{probe, \Pe}}}\xspace}
\newcommand{\Zjet}{\ensuremath{\PZ+\text{jet}}\xspace}
\newcommand{\ptGamLead}{\ensuremath{\pt^{\PGg_1}}\xspace}
\newcommand{\ptGamSublead}{\ensuremath{\pt^{\PGg_2}}\xspace}
\newcommand{\ptH}{\ensuremath{\pt^{\PH}}\xspace}
\newcommand{\absYH}{\ensuremath{\abs{y^\PH}}\xspace}
\newcommand{\NJ}{\ensuremath{N_{\text{Jets}}}\xspace}
\newcommand{\PTJEins}{\ensuremath{\pt^{\mathrm{j}_1}}\xspace}
\newcommand{\sigmaEff}{\ensuremath{\sigma_{\text{eff}}}\xspace}
\newcommand{\xH}{\ensuremath{\mathrm{x\PH}}\xspace}

\newcommand{\ECALLow}{\ensuremath{1.4442}\xspace}
\newcommand{\ECALHigh}{\ensuremath{1.5660}\xspace}

\cmsNoteHeader{HIG-23-014}
\title{Measurements of inclusive and differential Higgs boson production cross sections at \texorpdfstring{$\sqrt{s} = 13.6\TeV$}{sqrt(s) = 13.6 TeV} in the \texorpdfstring{$\PH\to\PGg\PGg$}{H to gamma gamma} decay channel}

\date{\today}

\abstract{
   Inclusive and differential cross sections for Higgs boson production in proton-proton collisions at a centre-of-mass energy of 13.6\TeV are measured using data collected with the CMS detector at the LHC in 2022, corresponding to an integrated luminosity of 34.7\fbinv. Events with the diphoton final state are selected, and the measured inclusive fiducial cross section is $\sigFid=74\pm11\stat^{+5}_{-4}\syst\fb$, in agreement with the standard model prediction of $67.8\pm3.8\fb$. Differential cross sections are measured as functions of several observables: the Higgs boson transverse momentum and rapidity, the number of associated jets, and the transverse momentum of the leading jet in the event. Within the uncertainties, the differential cross sections agree with the standard model predictions.
}

\hypersetup{
pdfauthor={CMS Collaboration},
pdftitle={Measurements of inclusive and differential Higgs boson production cross sections at sqrts = 13.6 TeV in the H to gamma gamma decay channel},
pdfsubject={CMS},
pdfkeywords={CMS, Higgs boson, Run 3, diphoton, cross section, differential}}

\maketitle

\section{Introduction}
\label{sec:intro}
The Higgs boson (\PH) was discovered in 2012 by the ATLAS~\cite{HiggsAtlas} and CMS~\cite{HiggsCMS, HiggsCMS1} Collaborations using proton-proton (pp) collisions at the CERN LHC~\cite{Evans:2008zzb}.
Since then, many properties of the Higgs boson have been examined~\cite{ATLAS:2022vkf, CMS:2022dwd}.
The large amount of data collected since the observation permits detailed testing of the standard model (SM) with differential measurements.
Whereas measurements in the simplified template cross section scheme target the different production modes and their differential properties, the fiducial inclusive and differential cross sections account for the sum of all production mode contributions to the fiducial phase space and are hence less model dependent~\cite{LHCHiggsCrossSectionWorkingGroup:2016ypw}.

The ATLAS and CMS Collaborations measured fiducial inclusive and differential cross sections at the centre-of-mass energies of 7, 8, and 13\TeV in the \Hgg~\cite{ATLAS:2014yga, CMS:2015qgt, ATLAS:2022fnp, CMS:2022wpo}, \HZZStarFourEll~\cite{ATLAS:2014xzb,CMS:2015zpx,ATLAS:2020wny,CMS:2023gjz}, \HWWStarEllPrimed~\cite{ATLAS:2016vlf, CMS:2016ipg,ATLAS:2023hyd,ATLAS:2023pwa,CMS:2020dvg}, and \Htautau~\cite{CMS:2021gxc,ATLAS:2024wfv} decay channels.
The measurements in the diphoton, four-lepton, and $\PW \PW$ decay channels by the ATLAS~\cite{ATLAS:2020wny,ATLAS:2022fnp,ATLAS:2022ooq} and CMS~\cite{CMS:2023gjz,CMS:2022wpo,CMS:2020dvg} Collaborations at 13\TeV reach a relative uncertainty of 8--11\%.
The combination of the ATLAS measurements in the \Hgg and \HZZStarFourEll decay channels under the SM assumptions results in an uncertainty of 7\% in the total cross section for Higgs boson production~\cite{ATLAS:2022qef}.

In 2022, the Run 3 of the LHC started at the increased centre-of-mass energy of 13.6\TeV, providing the opportunity to extend Higgs boson cross section measurements to the increased energy and further test the SM predictions.
Inclusive fiducial cross sections for Higgs boson production were measured at \sqrtstpstev in the \Hgg and \HZZStarFourEll decay channels by the ATLAS Collaboration with relative uncertainties of about 17 and 26\%, respectively~\cite{ATLAS:2023tnc}.
The CMS Collaboration measured the inclusive fiducial cross section in the \HZZStarFourEll decay channel at 13.6\TeV with a relative uncertainty of $\approx20\%$~\cite{CMS:2025wnr}.
In addition, differential fiducial cross sections were measured as a function of the transverse momentum (\pt) and absolute value of the rapidity of the Higgs boson, denoted by \ptH and \absYH, respectively.
This paper presents inclusive and differential measurements of the Higgs boson production in the \Hgg channel from the CMS experiment at \sqrtstpstev.
The measurements use data recorded with the CMS detector~\cite{CMS:2008xjf, CMS:2309.05466} in 2022 and are performed in a fiducial phase space defined at the particle level with improved perturbative convergence of the theoretical calculations~\cite{Salam:2021tbm}.

Although the \Hgg decay has a branching fraction ($\mathcal{B}$) of $\approx 0.227\%$~\cite{LHCHiggsCrossSectionWorkingGroup:2016ypw}, the excellent energy resolution of the CMS electromagnetic calorimeter (ECAL) yields a narrow peak in the invariant-mass distribution of the two photons, \mgg.
The analysis follows the strategy used in previous measurements in this channel~\cite{CMS:2015qgt, CMS:2018ctp, CMS:2022wpo}.
Events are selected with two reconstructed photons with \mgg $\in [100,180]\GeV$, \ie in a range around the Higgs boson mass of $\approx 125\GeV$.
The background and signal contributions are then estimated from a combined fit of a smoothly falling and a peaking function to the \mgg distribution, reducing the reliance on Monte Carlo (MC) simulations for background modelling.
The background function and its parameters are determined from the fit to data.
The signal function is estimated from MC simulations corrected using \Zee decays, taking advantage of the similarity of electron and photon electromagnetic showers.
To improve the measurement sensitivity, events are categorized based on a per-event \mgg resolution estimate.
A combined fit to all categories is performed to extract the number of signal events.
Compared to the previous measurement of fiducial inclusive and differential cross sections by the CMS Collaboration, which used 137\fbinv of pp collision data at \sqrtsttev~\cite{CMS:2022wpo}, this analysis contains several improvements.
In particular, a novel approach based on a neural network is used to correct the modelling of photon identification variables and the estimate of the per-photon energy resolution in MC simulations, so that their distributions agree better with those observed in data.

The paper is structured as follows:
the CMS detector is briefly introduced in Section~\ref{sec:cms}.
The data and simulation samples are described in Section~\ref{sec:samples}.
The event reconstruction, and the event selection and categorization are presented in Sections~\ref{sec:event_reco} and~\ref{sec:selection_categorization}, respectively.
The fiducial phase space and the observables for the differential measurements are introduced in Section~\ref{sec:psDef_observables}.
The statistical analysis and the systematic uncertainties are described in Sections~\ref{sec:statistical_analysis} and~\ref{sec:systematic_uncertainties}, respectively.
The results are presented in Section~\ref{sec:results}, followed by a summary in Section~\ref{sec:summary}.
Tabulated results are provided in the HEPData record for this analysis~\cite{hepdata}.

\section{The CMS detector}
\label{sec:cms}
The central feature of the CMS apparatus is a superconducting solenoid of 6\unit{m} internal diameter, providing a magnetic field of 3.8\unit{T}.
Within the solenoid volume are a silicon pixel and strip tracker, a lead tungstate crystal ECAL, and a brass and scintillator hadron calorimeter, each composed of a barrel and two endcap sections.
The ECAL consists of 75\,848 lead tungstate crystals, which provide coverage in pseudorapidity $\abs{\eta} < 1.48 $ in the barrel region (EB) and $1.48 < \abs{\eta} < 3.0$ in the two endcap regions (EE).
Preshower detectors consisting of two planes of silicon sensors interleaved with a total of three radiation lengths of lead are located in front of each EE detector.
Forward calorimeters extend the pseudorapidity coverage provided by the barrel and endcap detectors.
Muons are reconstructed using gas-ionization detectors embedded in the steel flux-return yoke outside the solenoid.

Events of interest are selected using a two-tiered trigger system.
The first level, composed of custom hardware processors, uses information from the calorimeters and muon detectors to select events at a rate of around 100\unit{kHz} within a fixed latency of about 4\mus~\cite{CMS:2020cmk}.
The second level, known as the high-level trigger (HLT), consists of a farm of processors running a version of the full event reconstruction software optimized for fast processing, and reduces the event rate to around 5\unit{kHz} before data storage~\cite{CMS:2016ngn,CMS:2024aqx}.

A more detailed description of the CMS detector, together with a definition of the coordinate system used and the relevant kinematic variables, can be found in Refs.~\cite{CMS:2008xjf,CMS:2309.05466}.

\section{Data samples and simulated events}
\label{sec:samples}
This analysis uses pp collision data, collected in 2022 at \sqrtstpstev, corresponding to an integrated luminosity of 34.7\fbinv~\cite{LUM-22-001}.
The data were selected using a diphoton HLT~\cite{CMS:2016ngn} with \pt thresholds on the highest \pt (leading, \gamOne) and second-highest \pt (subleading, \gamTwo) photon of 30 and 22\GeV, respectively.
Additionally, it is required that $\mgg > 90\GeV$.
Both photons were required to pass loose identification criteria on the amount of surrounding energy deposits in the calorimeters and on variables that characterize the photon shower in the ECAL (shower shape variables)~\cite{CMS:2020uim}.

The four main Higgs boson production modes are gluon-gluon fusion (\ggH), vector boson fusion (\VBF), associated production with a W or Z boson (\VH), and with a top quark pair (\ttH).
Samples for these processes are generated with \MGvATNLO (version 2.9.9)~\cite{AMCAT} at next-to-leading order (NLO) in perturbative quantum chromodynamics using the NNPDF3.1 NNLO~\cite{NNPDF:2017mvq} set of parton distribution functions (PDFs).
The FxFx merging scheme~\cite{Frederix:2012ps} is used to match jets from matrix element calculations to those from parton shower for the \ggH, \VH, and \ttH production processes.
Events in the ggH production mode are weighted to match the predictions from the \NNLOPS generator~\cite{NNLOPS1, NNLOPS2, NNLOPS3} as a function of the Higgs boson~\pt and the number of jets in the event.
The samples are normalized to the cross sections provided by the LHC Higgs Working Group for \sqrtstpstev~\cite{13p6_LHCHWG_extrap}, based on an interpolation procedure using cross sections computed for the centre-of-mass energies of 13 and 14\TeV~\cite{LHCHiggsCrossSectionWorkingGroup:2016ypw,13p6_LHCHWG_extrap}, using the results from Refs.~\cite{Anastasiou:2015ema,Anastasiou2016,Ciccolini:2007jr,Ciccolini:2007ec,Bolzoni:2010xr,Bolzoni:2011cu,Brein:2003wg,Ciccolini:2003jy,Beenakker:2001rj,Beenakker:2002nc,Dawson:2002tg,Dawson:2003zu,Yu:2014cka,Frixione:2014qaa,Demartin:2015uha,Demartin:2016axk,Denner:2011mq,Djouadi:2018xqq,Bredenstein:2006rh,Bredenstein:2006ha,Boselli:2015aha,Actis:2008ts}.

The main backgrounds are the irreducible background from non-resonant diphoton (\gamgam) production and the reducible background from \gjet production, where a jet is misreconstructed as a photon.
Multijet production accounts for an additional and smaller contribution in the reducible background.
For the measurement of the inclusive and differential cross sections, the normalization and the shape of the \mgg background distribution are estimated from the data.
Simulated events for \gamgam and \gjet production are only used to optimize the event selection and categorization.
Diphoton production is simulated with \SHERPA (version 2.2.12)~\cite{SHERPA22} at leading order (LO).
The gluon-induced box process is simulated without additional final-state partons in the matrix element, whereas the quark-induced and quark-gluon-induced Born processes include up to three additional partons in the final state.
The production of \gjet events is simulated with {\PYTHIA}8~\cite{Bierlich:2022pfr} (version 8.306) as a $2\to2$ LO process at the matrix-element level.

Several corrections are applied to the simulation samples or to the data to either assess and address mismodelling in the simulation or to improve the data calibration.
These corrections are inferred from a comparison of data and simulation in events with \Zee and \Zmmg decays and are validated with these samples.
In data, \Zee events are collected using a single-electron trigger with a \pt\ threshold of 30\GeV and a double-electron trigger with \pt thresholds of 23 and 12\GeV for the leading and subleading electron, respectively~\cite{CMS:2016ngn,CMS:2020uim,CMS:2024aqx}.
Events with a radiative \Zmmg decay are recorded with dimuon triggers with \pt thresholds of 17 and 8\GeV for the leading and subleading muon, respectively~\cite{CMS:2016ngn,CMS:2021yvr,CMS:2024aqx}.
The simulation samples used for these corrections are generated with \MGvATNLO.

All simulated events are interfaced with \PYTHIA for the simulation of the parton shower, fragmentation, and hadronization, using the \textsc{CP5} underlying event tune~\cite{CP5}.
The \PYTHIA generator is also used to simulate additional pp interactions occurring in the same or neighbouring bunch crossings (pileup).
Simulated events are reweighted to reproduce the distribution of the number of interaction vertices observed in the data.
The average number of interactions per bunch crossing in 2022 data is 46, assuming a total inelastic pp cross section of 80\mb at \sqrtstpstev.
For all processes, the response of the CMS detector is simulated using a detailed description of the CMS apparatus based on the \GEANTfour package~\cite{GEANT4:2002zbu}.

\section{Event reconstruction}
\label{sec:event_reco}
The primary vertex is taken as the vertex corresponding to the hardest scattering in the event, which is defined as the vertex that maximizes the $\pt^2$ sum of reconstructed particles, evaluated using tracking information alone~\cite{CMS-TDR-15-02}.
A particle-flow (PF) algorithm~\cite{CMS:2017yfk} aims to reconstruct and identify each individual particle in an event (PF candidate), with an optimized combination of information from the various elements of the CMS detector.

Following a power cooling issue in September 2022, about 7\% of the ECAL channels in one of the EEs were disabled~\cite{CMS-DP-2024-052}.
Events with at least one jet with $\pt > 30\GeV$ in this region are removed to prevent biases that could affect the cross section measurements reported in this paper.
Simulation samples are split into two periods to reflect the detector conditions before and after the issue and are weighted according to the integrated luminosities of the corresponding data sets.

\subsection{Photon reconstruction and identification}
\label{subsec:photons}

Energy deposits (clusters) in the ECAL form the basis of the photon reconstruction~\cite{CMS:2020uim}.
The ECAL clusters with energy well above the electronics noise level are combined if they are compatible with originating from the same photon, resulting in a so-called supercluster.
Superclusters are selected as photons not matched to charged-particle trajectories associated with a reconstructed hard-scattering vertex.
Photons in the transition regions between the barrel and the endcaps ($\ECALLow < \abs{\eta} < \ECALHigh$) are not considered.

An energy resolution of $\approx1\%$ is achieved in the EB for unconverted or late-converting photons in the tens of \GeV energy range.
The energy resolution for other photons is $\approx$1.3\% up to $\abs{\eta} = 1$ and $\approx2.5\%$ up to $\abs{\eta} = 1.44$ in the EB.
In the EE, the energy resolution is $\approx2.5\%$ for unconverted or late-converting photons, and between 3 and 4\% for other photons.

Not all the photon energy is deposited in the operational ECAL crystals or accounted for in the clustering.
This is mainly due to lateral and longitudinal shower leakage, intermodule gaps, unresponsive channels, or energy thresholds used to mitigate the noise in the cluster reconstruction.
To improve the energy measurement, a dedicated simulation-based parametric regression is used.
This method also provides a per-photon estimate for the energy resolution, $\sigma_E$.
The input variables to the regression include the location of the supercluster and its seed (the crystal with the highest energy), and a range of shower shape variables.
The relative energy resolution of photons with generated transverse momenta in the range of 1 to 100\GeV and with $\abs{\eta} < 1.0$ improves from about 1.3 to 0.9\% when the regression is applied.
In the outer EE ($2.0 < \abs{\eta} < 2.5$), the relative resolution for photons in the same momentum range improves from approximately 3.1 to 2.4\%.

Differences between data and simulation in the energy scale and resolution are addressed by residual corrections.
Electrons from \Zee decays that are reconstructed and calibrated with the photon algorithms described above, except for the track-based veto, are used.
The corrections are derived from a comparison of the dielectron invariant mass distribution in data and simulation~\cite{CMS:2020xrn}.
The scale calibrations are applied to data and are derived from the positions of the \PZ boson peak in data and simulation as functions of time, the gain from the photodetector readout of the supercluster's seed crystal, as obtained from its multigain preamplifier~\cite{CMS:1997ysd}, the photon energy, the supercluster $\eta$, and the $\RNINE$ variable.
The $\RNINE$ variable is defined as the energy deposited in the three-by-three crystal matrix around the seed divided by the total uncorrected energy of the supercluster~\cite{CMS:2020uim} and is useful to discriminate converted from unconverted photons.
The resolution corrections broaden the width of the distribution in simulation, such that it corresponds to that observed in the data, and depend on the photon energy, the supercluster $\eta$, and $\RNINE$.

\begin{figure}
    \centering
        \includegraphics[width=0.55\textwidth]{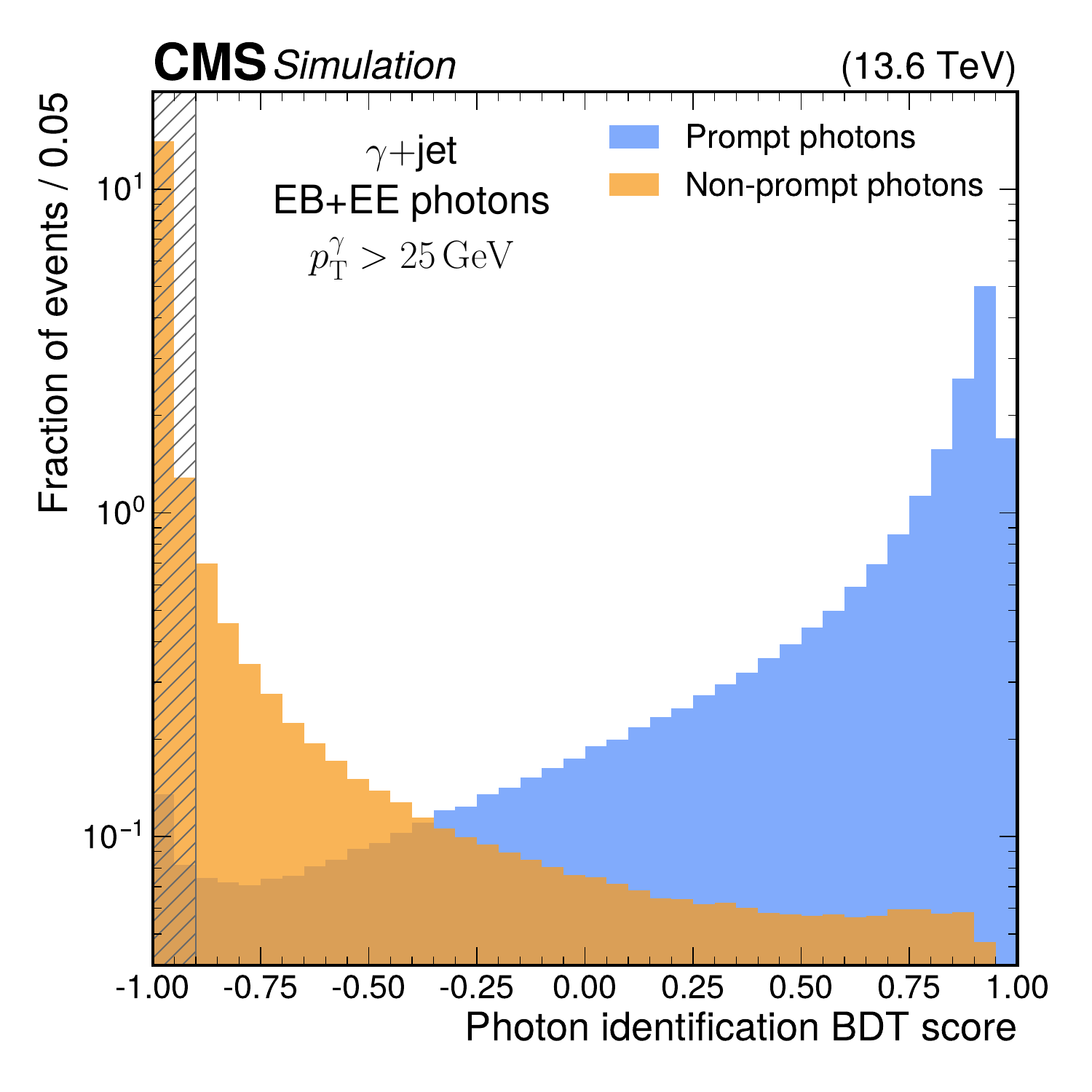}
    \caption{Normalized distributions of the photon identification BDT scores for prompt (blue) and non-prompt (orange) photons from \gjet simulated events.
    The shaded region indicates the photons that are rejected by the photon preselection requirement of ${>} -0.9$.}
    \label{fig:idmva_distributions}
  \end{figure}

A boosted decision tree (BDT) is trained to define photon identification (ID) criteria~\cite{CMS-DP-2024-052} in a similar way as described in Ref.~\cite{CMS:2020uim}.
The BDT score is used to separate prompt photons that are produced at the primary vertex and non-prompt photons, which mostly stem from collimated diphoton decays of neutral mesons, such as \PGpz and \Pgh, inside hadronic jets.
As input, the model uses shower shape variables, isolation variables built from the \pt of tracks and PF objects near the photon candidate, and the ratio of the energy deposited in the hadronic calorimeter behind the photon supercluster to the supercluster energy (\HoE).
The BDT is implemented in \textsc{XGBoost}~\cite{xgboost} and is trained on prompt and non-prompt photons from simulated \gjet events.
Besides the discriminant features, the energy and $\eta$ of the photon supercluster as well as $\rho$, defined as the median of the transverse energy density in the event, are also used as input features.
The BDT scores of prompt and non-prompt photons from simulated \gjet events are shown in Fig.~\ref{fig:idmva_distributions} for reconstructed photons with $\ptGam > 25\GeV$.
The leading and the subleading photon are included in the distributions.
Prompt photons accumulate at high values of the BDT score, whereas the distribution of non-prompt photons is steeply falling towards larger values.
The photon preselection used in this analysis rejects photons with a BDT score smaller than $-0.9$, as indicated by the shaded region.
This requirement is highly efficient for selecting prompt photons and rejects a large fraction of non-prompt photons, reducing the number of possible diphoton pairs.

Electron background contributions are reduced with a conversion-safe electron veto algorithm~\cite{CMS:2015myp}.
This veto rejects the photon candidate if its supercluster is close to a track compatible with an electron, unless the track is matched to a photon conversion vertex.

\subsection{Corrections to simulated photons}
\label{sec:corr}

Photon mismodelling in the simulation is a non-negligible source of systematic uncertainty in \Hgg fiducial cross section measurements.
This holds in particular for shower shape and isolation variables~\cite{CMS:2015qgt}, but also for the per-photon energy resolution estimate,~$\sigma_E$.
Sources of the mismodelling include the imperfect description of the material budget in the simulation, the modelling of noise, and the time evolution of the detector response, in particular due to radiation damage~\cite{CMS:2020uim,CMS:2022wpo}.
Shower shape and isolation variables, as well as $\sigma_E$, are hence corrected using probe electrons reconstructed as photons (\gamProbee) with the tag-and-probe method~\cite{TAG-PROBE} in \Zee events.
A first correction method, called ``chained quantile-regression'', was developed in Ref.~\cite{CMS:2022wpo} for the analysis of the Run 2 data.
It involves the training of a large number of BDTs and was used to successfully correct shower shape and isolation variables.

In this analysis, a new method based on normalizing flows~\cite{papamakarios2021normalizing_review} is used for a more efficient derivation of the corrections.
It takes as input the variables that are used for the photon identification BDT, as well as the per-photon estimate of the energy resolution, and provides as output per-photon corrections for all of these variables.
Hence, the method does not only correct these values but also their correlations.
The basis of the method is that normalizing flows are able to learn a high-dimensional mapping from a distribution of interest, in our case the BDT input variables and $\sigma_E$, to a simpler distribution of same dimension, often a multivariate Gaussian.
We use a simple but efficient solution with one normalizing flow~\cite{Daumann:2024kfd} trained on both data and simulation.
It is parametrized as a function of an MC/data binary variable, as well as the \pt, $\eta$, and the azimuthal angle $\phi$ of the \gamProbee, and $\rho$.
To perform the correction, the mapping to the multivariate Gaussian is performed for simulated photons, the MC/data boolean is flipped, and the inverse transformation provides the corrected values.
The model is an autoregressive normalizing flow~\cite{MAF} and consists of five neural spline transformations~\cite{Spline_flows} with ten spline bins, implemented using the \textsc{PyTorch}~\cite{pytorch} and \textsc{Zuko}~\cite{zuko} libraries.

Before deriving these corrections, the distributions of the conditional variables $\rho$, \pt, $\eta$, and $\phi$ are reweighted in simulation to match the data distributions.
In addition, the isolation variables, which show a discontinuous behaviour with a peak at zero followed by a continuous tail due to \pt thresholds on PF candidates and energy depositions in the calorimeters, are transformed to a continuous distribution~\cite{Daumann:2024kfd}: isolation values in the peak are resampled to populate the gap between the peak and the start of the continuous tail.

\begin{figure}
    \centering
        \includegraphics[width=0.4\textwidth]{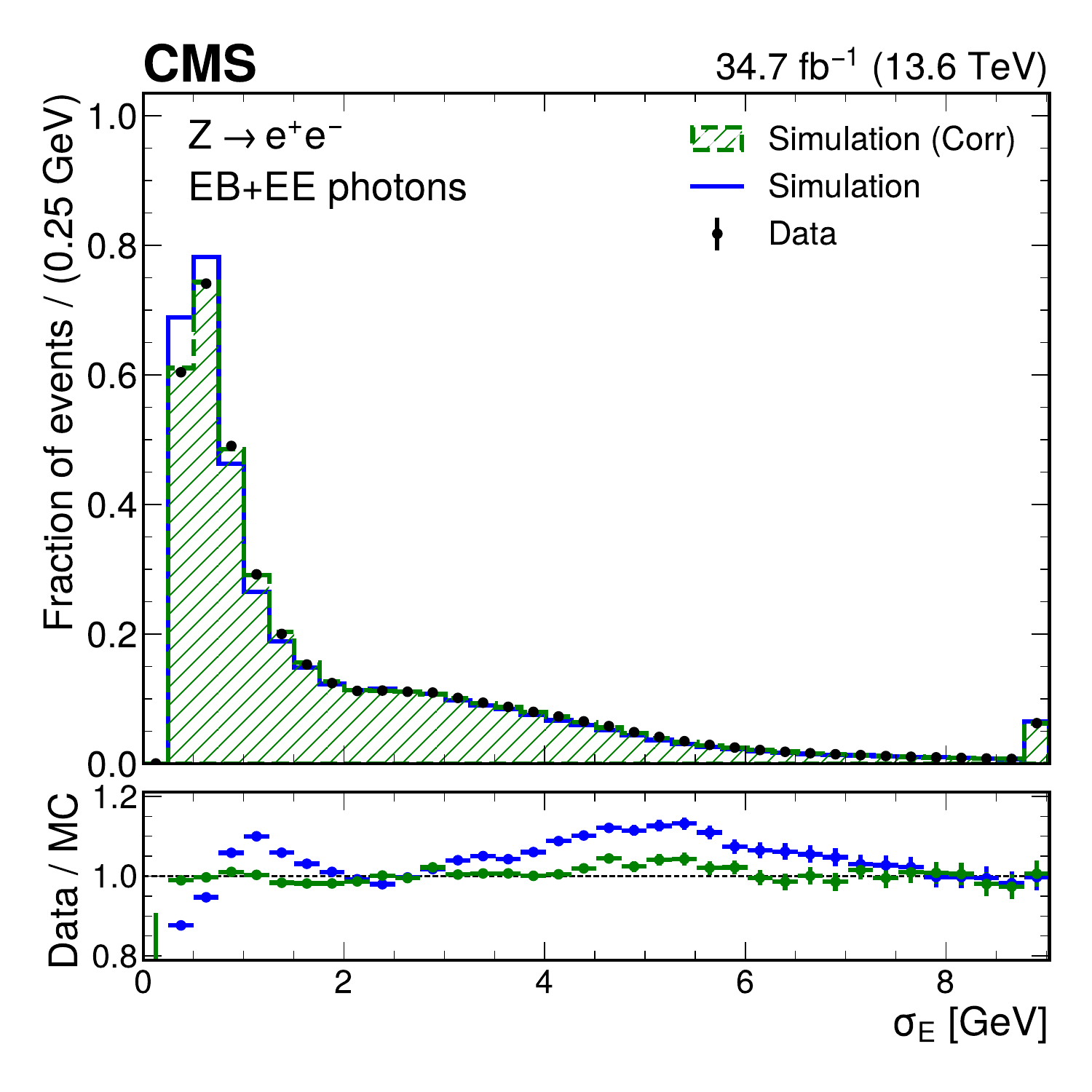}
        \includegraphics[width=0.4\textwidth]{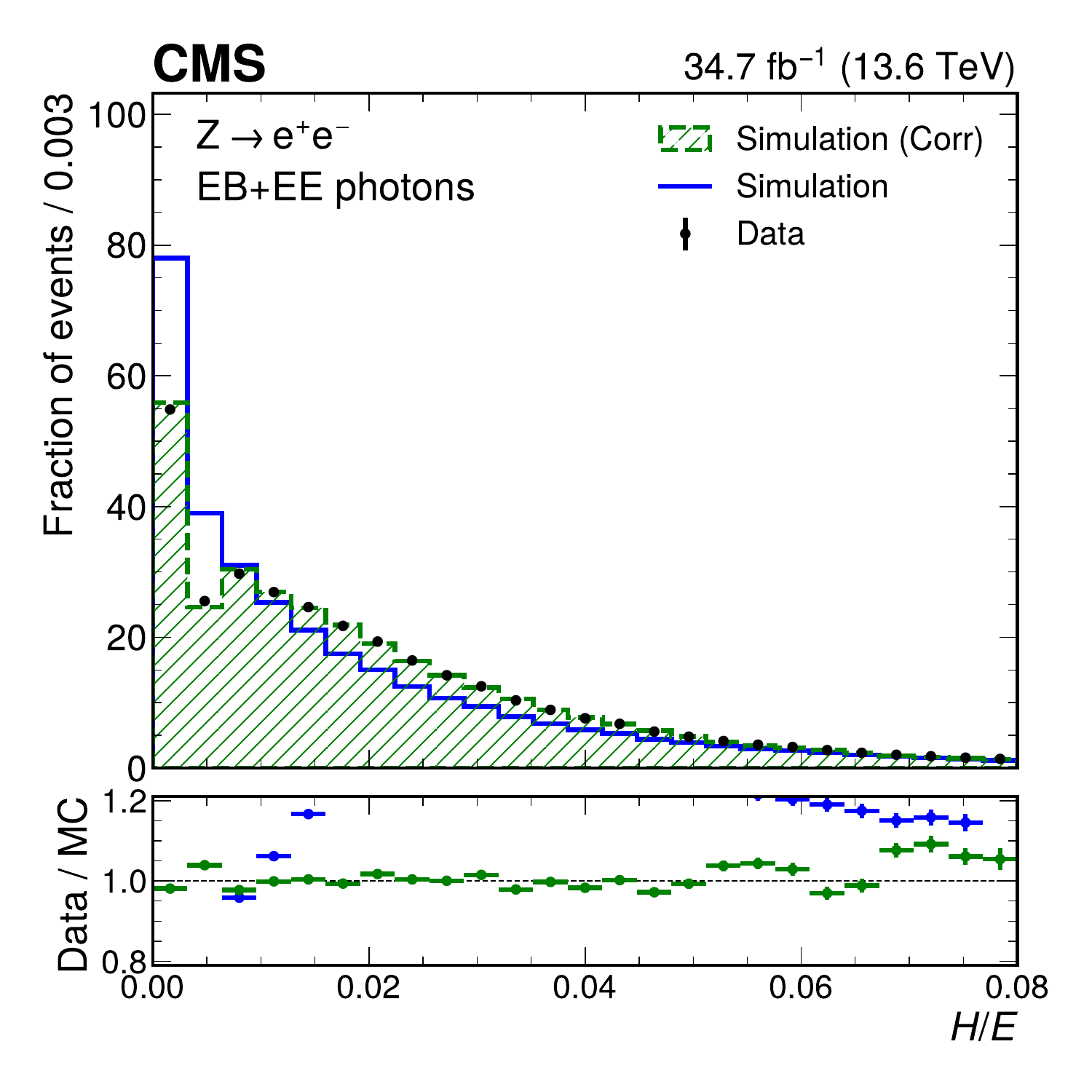}
        
        \includegraphics[width=0.4\textwidth]{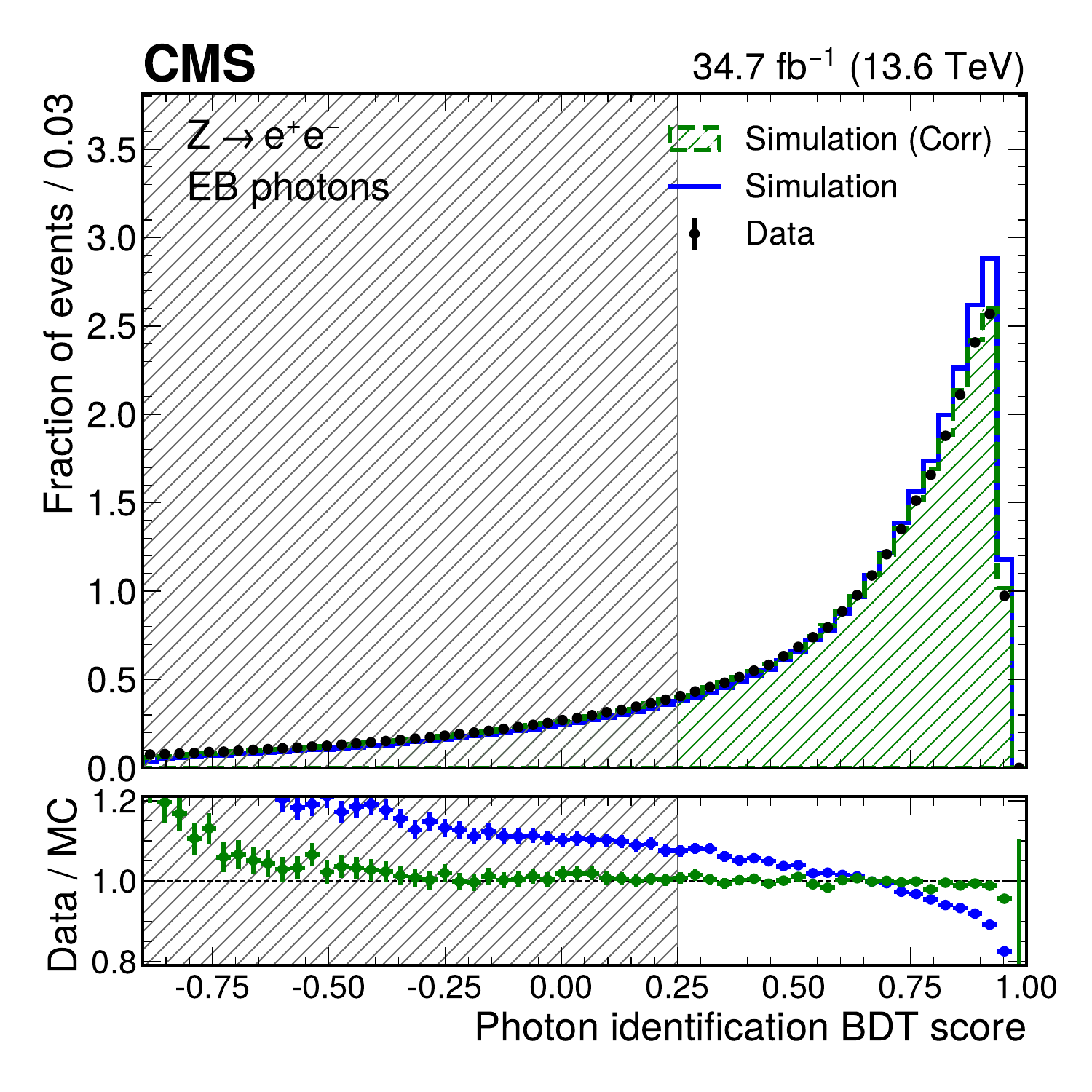}
        \includegraphics[width=0.4\textwidth]{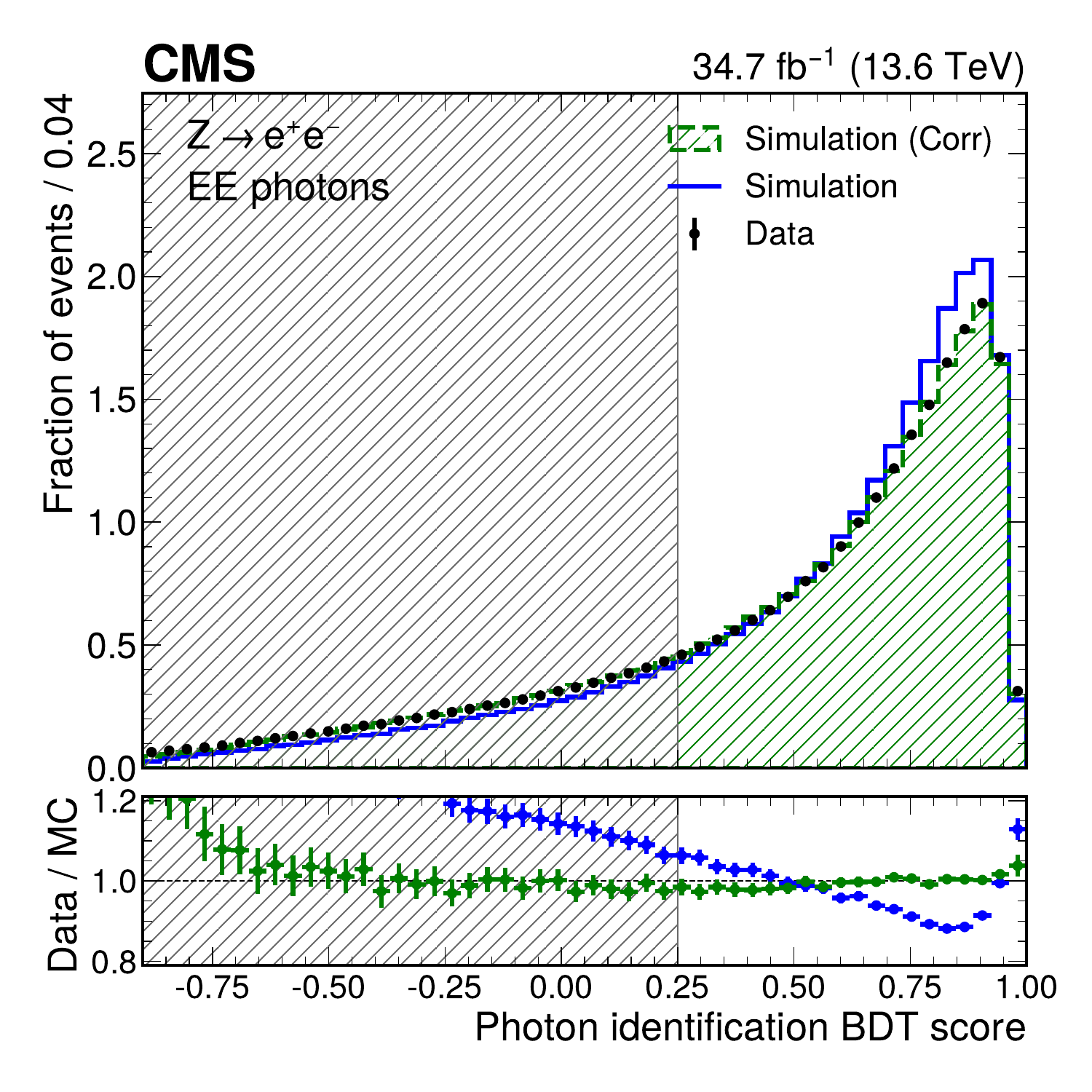}
        \caption{
        Data-to-simulation comparison for $\sigma_{E}$ (upper left), \HoE (upper right), the photon identification BDT score in EB (lower left) and EE (lower right) for electrons from \Zee decays reconstructed as photons.
        The uncorrected distributions are shown in blue and the corrected distributions from the normalizing flow are shown in green.
        The error bars in the ratio panels include the statistical uncertainty from the data and the uncertainty from the limited number of simulated events.
        For the distributions of the photon identification BDT score, the shaded region corresponds to photons with a BDT score ${<}0.25$, which are excluded by the selection applied in the cross section measurements.
        For the $\sigma_{E}$ distribution, the last bin contains the overflow.
        }
        \label{fig:photon_ID_BDT_output}
\end{figure}

The level of agreement between selected distributions in data and simulation for \gamProbee from \Zee decays before and after the corrections is shown in Fig.~\ref{fig:photon_ID_BDT_output}.
The distributions for $\sigma_E$ and for \HoE are shown, where \HoE serves as an example for the input variables that are used in the identification BDT with the largest observed data-to-simulation shape disagreement between data and MC.
In addition, the BDT score is shown separately for photons in the EB and in the EE.
After the corrections, the agreement of the BDT score between data and simulation is significantly improved.
The average disagreement is as small as 1.7 (2.0)\% in the EB (EE), where the average is calculated from the absolute difference between data and MC in all bins shown in Fig.~\ref{fig:photon_ID_BDT_output} (lower row).
The corrections are validated using photons from \Zmmg events.
Data and simulations are found to agree within the uncertainties for the BDT score and $\sigma_E$ variables, which enter the event selection and categorization.

\subsection{Lepton reconstruction}
\label{subsec:leptons}

Electrons and muons are used for the derivation and validation of corrections in \Zee and \Zmmg events and for overlap removal procedures to resolve ambiguities between reconstructed objects.
The reconstruction of electrons is based on charged-particle tracks matched to ECAL clusters~\cite{CMS:2020uim}.
Superclusters are built in the same way as for photons.
The electron momentum is estimated from a combination of the ECAL energy and the momentum measurement in the tracker.
Electrons are required to have $\pt > 15\GeV$ and have to satisfy loose cut-based identification requirements with a signal efficiency of $\approx90\%$.
The variables used to define these identification criteria are described in Ref.~\cite{CMS:2020uim}.
Muons are reconstructed using a combination of a track in the central tracking system and either a single track or multiple hits in the muon detectors~\cite{CMS:2018rym}.
Muons are required to have $\pt > 10\GeV$ and they need to pass tight identification criteria, which are based on the relative isolation with respect to hadrons and photons in a cone of radius $\DR=0.4$, where $\DR=\sqrt{(\Delta\eta)^{2}+(\Delta\phi)^{2}}$ defines the angular distance, around the muon and the quality of the fit of the muon track.
Electrons and muons have to be separated by an angular distance greater than 0.2 from both of the photons of the diphoton pair, which is reconstructed as described in Section~\ref{subsec:event_selection}, when they are used for overlap removal.
No requirement on the angular distance to photons is placed when electrons and muons are used for the derivation of corrections in \Zee and \Zmmg events.

\subsection{Jet reconstruction}
\label{subsec:jets}

In this analysis, jets are only used for the differential measurement with respect to the number of jets.
Jets are reconstructed from PF candidates using the anti-\kt algorithm~\cite{antikt,FastJet} with a distance parameter of 0.4.
The pileup-per-particle identification algorithm~\cite{CMS:2020ebo} is used to mitigate pileup effects.
The algorithm assigns a weight to each particle candidate before the jet clustering according to the likelihood that the candidate originated from pileup.
Jets originating from noise and reconstruction failures are rejected using criteria on the energy composition and number of PF constituents of the jets~\cite{CMS:2020ebo}.

Jet energy corrections (JECs) are derived from simulation to calibrate the measured jet momentum to that of particle-level jets~\cite{CMS:2016lmd}.
In situ measurements in data of the momentum balance in dijet, \gjet, \Zjet, and multijet events are used to account for residual differences in the jet energy scale between data and simulation.
The jet energy resolution (JER) is found to be worse in data than in simulation.
The resolution in the simulation is hence broadened to agree with that observed in data.

Jets are required to have $\pt > 30\GeV$ and $\abs{\eta}<2.5$, as jets in the central part of the detector are subject to lower systematic uncertainties than in the forward region.
Whereas the requirement on $\abs{\eta}$ reduces the efficiency for the VBF process, inclusive Higgs boson production is dominated by the ggH process in the SM.
Jets with $\DR < 0.4$ from a photon of the diphoton pair or a charged lepton are removed.

\section{Event selection and categorization}
\label{sec:selection_categorization}
\subsection{Event selection}
\label{subsec:event_selection}
The event selection retains \Hgg signal candidates while rejecting as much background as possible from both prompt and non-prompt photons.
Each photon must have a supercluster with $\abs{\eta}< 2.5$, excluding the ECAL barrel-endcap transition regions of $\ECALLow < \abs{\eta} < \ECALHigh$.
Each photon must satisfy preselection criteria based on its shower shape, isolation and kinematic properties.
These criteria are described in detail in Ref.~\cite{CMS:2018piu} and they are defined to be slightly more stringent than the corresponding trigger requirements.
In particular, the requirements $\ptGamLead > 35\GeV$ and $\ptGamSublead > 25\GeV$ are applied to the $\pt$-leading and -subleading photons, respectively.
Only the $\pt$-leading diphoton system composed of photons satisfying the preselection requirements is considered.
The selection criteria for $\pt / \mgg$ are changed with respect to previous measurements~\cite{CMS:2021kom,CMS:2022wpo}.
Previously, the leading (subleading) photon had to fulfil $\pt / \mgg >  1/3$~($1/4$).
In this analysis, the requirement for the subleading photon is unchanged, but a requirement of $\sqrt{\smash[b]{\ptGamLead \ptGamSublead}} / \mgg > 1/3$ is applied instead of the above criterion for the leading photon.
This requirement on the scaled geometric mean of \ptGamLead and \ptGamSublead improves the perturbative convergence of the theoretical calculations~\cite{Salam:2021tbm}.

The efficiencies of the diphoton trigger are measured with \Zee events using the tag-and-probe method, from which corrections (``scale factors'') are derived for simulated events such that the efficiencies in simulation match those measured in the data~\cite{CMS:2016ngn}.
The measurement is performed in bins of \pt, $\eta$, and $\RNINE$.
Scale factors for the preselection efficiencies and for the photon identification efficiency after the application of the normalizing flow correction are measured in a similar way, whereas the scale factors for the efficiencies of the electron veto criterion are computed from \Zmmg events.

\subsection{Event categorization}
\label{subsec:categorisation}

In order to maximize the sensitivity to the Higgs boson signal and minimize the dependence on the underlying model for its production and decay, the selected events are categorized based on $\sigma_m / m$~\cite{CMS:2015qgt}, \ie the estimator of the per-event diphoton invariant-mass resolution divided by \mgg.
The resolution estimator $\sigma_m$ is calculated from the per-photon resolution estimator $\sigma_E$ obtained from the energy regression BDT (cf.~Section~\ref{subsec:photons}) for each photon:
\begin{equation}
    \frac{\sigma_m}{m} = \frac{1}{2} \sqrt{ \left( \frac{\sigma_{E_1}}{E_1} \right)^2 + \left( \frac{\sigma_{E_2}}{E_2} \right)^2  },
\end{equation}
where the contribution from the photon angles is neglected.
The contribution to the mass resolution from the angular resolution is negligible with respect to the one from the energy resolution if the chosen Higgs boson decay vertex has $\abs{\Delta z} < 1\unit{cm}$, with $\Delta z$ being the difference of the $z$-coordinate between the true and the reconstructed vertex.
This is referred to as the correct identification of the vertex, which occurs for $\approx70\%$ of ggH events.
The efficiency of the correct vertex identification increases with higher Higgs boson \pt and exceeds 85~(95)\% for VH (\ttH) associated production across the entire range.

Since the BDT is trained on simulation and the energy resolution is known to be worse in data than in simulation, the term that adjusts the energy resolution in simulation (cf.~Section~\ref{subsec:photons}) is added in quadrature to the value of $\sigma_E$ from the energy regression BDT before $\sigma_m / m$ is calculated.

The diphoton mass resolution in \Hgg decays is typically 1--2\%, depending on the measurement of the photon energies in the ECAL and the topology of the photons in the event~\cite{CMS:2020xrn}.
As the relative energy resolution $\sigma_E / E$ improves with photon energy, $\sigma_m / m$ is correlated with the diphoton invariant mass.
Hence, a categorization in $\sigma_m / m$ can distort the \mgg distribution of the background by depleting the low-mass region in high-resolution categories.
The background distribution, instead, is assumed to be monotonically falling over the selected diphoton mass range in the statistical analysis.
To avoid distortions of the invariant mass distribution, the mass-resolution estimator is decorrelated from the diphoton invariant mass.
This is achieved with the quantile morphing algorithm described in Ref.~\cite{CMS:2022wpo}.
It divides the $\sigma_m / m$ distribution into bins of \mgg and morphs the cumulative distribution function of $\sigma_m / m$ in each bin to resemble that in a reference mass bin, chosen as $[125.0, 125.5] \GeV$.

The decorrelated mass-resolution estimator is used to categorize the events.
Additionally, a requirement on the photon ID BDT score is placed on both photons to reject background contributions with non-prompt photons.
This requirement and the category boundaries are optimized simultaneously using simulated events.
The signal component is modelled with a combination of the four main production modes with a Higgs boson mass of 125\GeV, whereas the non-resonant background is composed of the diphoton and \gjet MC samples.
Both the signal and background components are weighted according to the expected number of events in the selected phase space, assuming the SM cross sections for the Higgs boson production processes at \sqrtstpstev~\cite{13p6_LHCHWG_extrap} and the LO cross section predictions for the background processes.
Additional scaling factors for the normalizations of the diphoton and \gjet processes are determined from a fit to the distribution of the lower of the two photon ID BDT scores, restricted to events where the minimum score exceeds zero, in order to improve the modelling of the invariant mass distribution.
The \mgg distribution of the signal and background samples is fitted using the sum of an exponential function for the background and the sum of two Gaussian distributions for the signal.
The figure of merit for the optimization is the approximate expected signal significance $S/\sqrt{B}$, where $S$ and $B$ are the expected number of signal and background events, respectively, in the smallest \mgg interval containing 68\% of simulated signal events.

\begin{figure}
    \centering
        \includegraphics[width=0.45\textwidth]{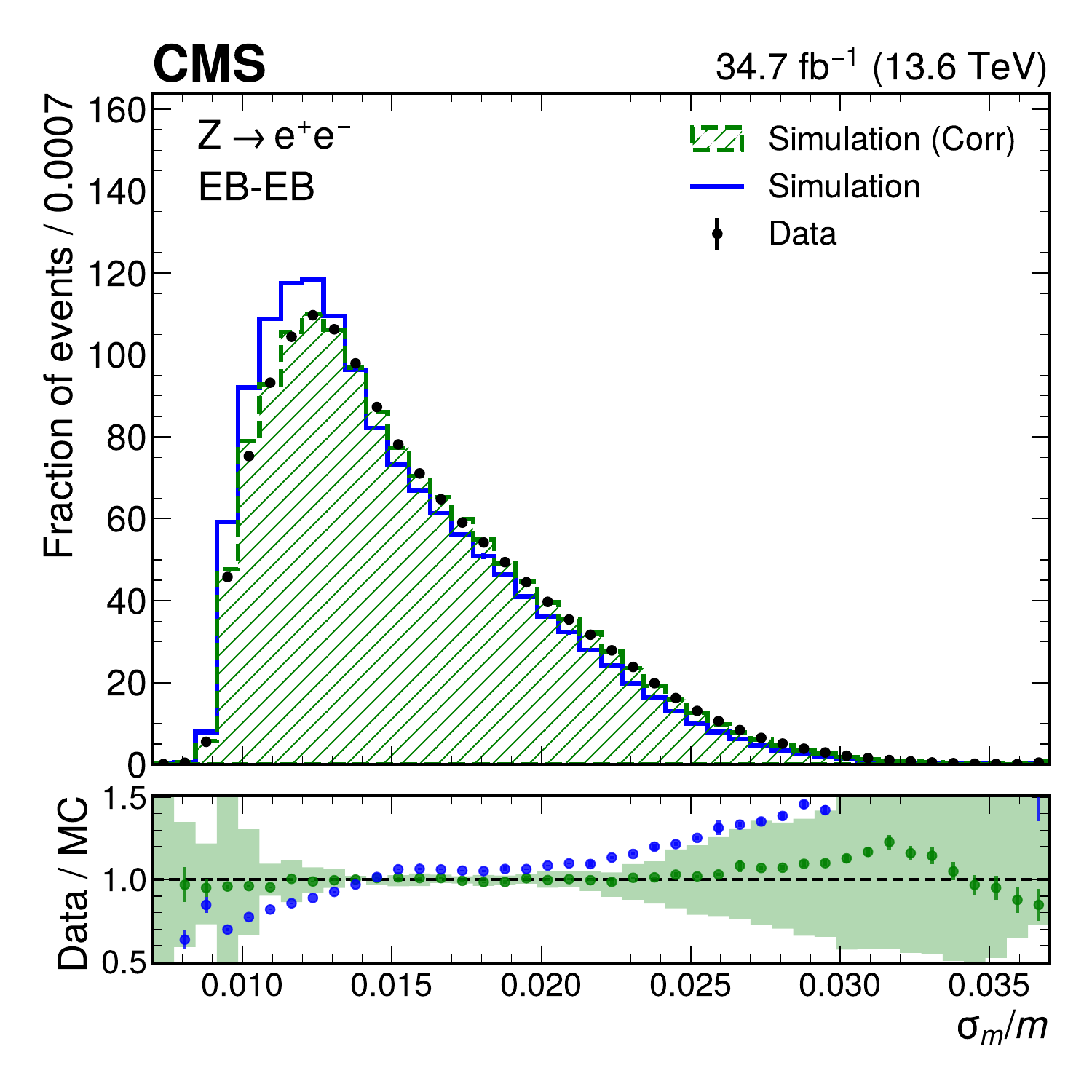}
        \includegraphics[width=0.45\textwidth]{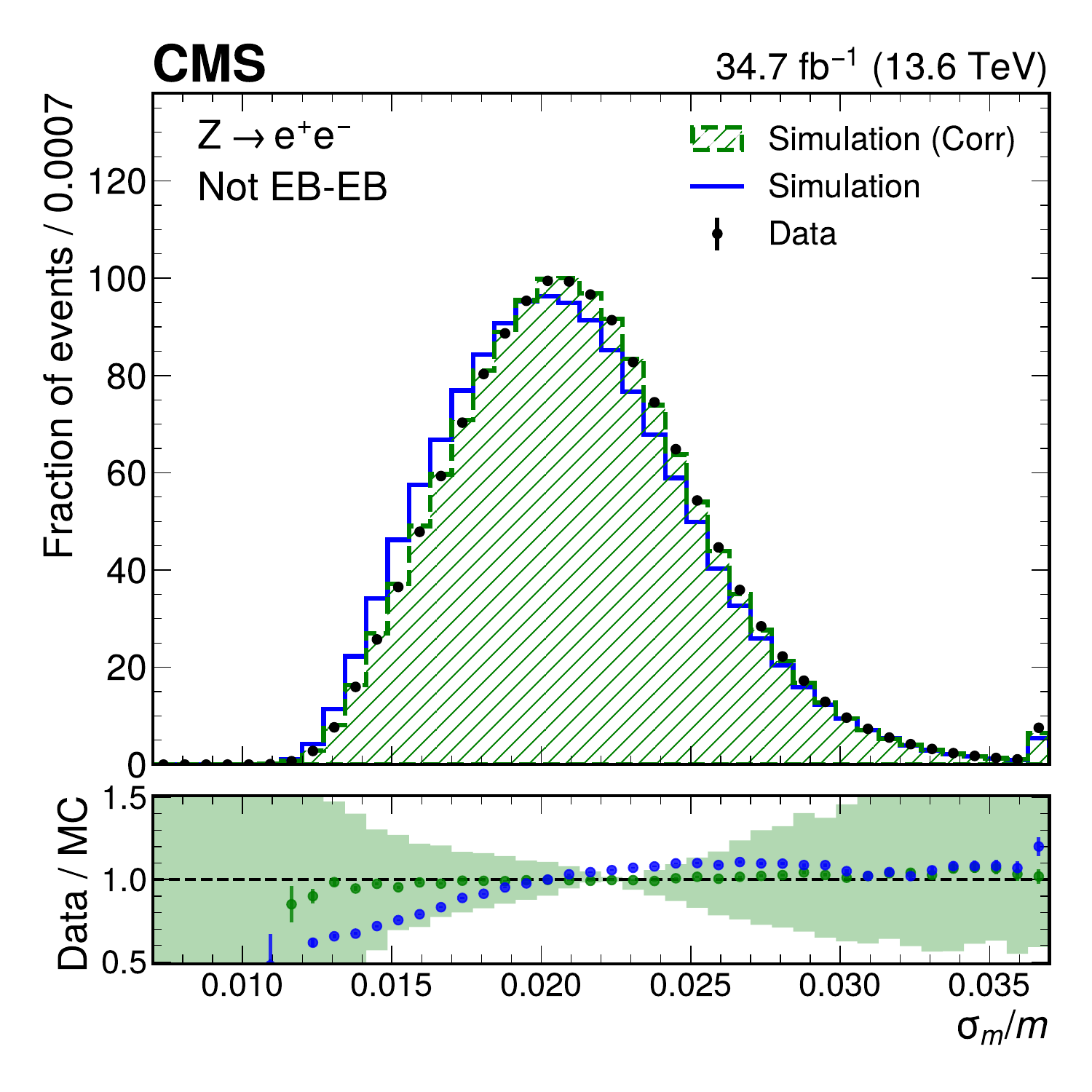}
        \caption{
        Data-to-simulation comparison of the per-event decorrelated mass-resolution estimator $\sigma_m/m$ using \Zee events.
        Both electrons are reconstructed as photons and categorized either both in the EB (left) or at least one in the EE (right).
        The uncertainty band in the lower panel represents the systematic uncertainty based on the residual mismodelling of $\sigma_E/E$ (5\%).
        The error bars on the markers in the lower panels include the statistical uncertainty from data and the uncertainty from a limited number of simulated events.
        The last bin contains the overflow.
        }
        \label{fig:sigma_m_over_m_decorrelated}
\end{figure}

The modelling of the $\sigma_m / m$ distribution in simulation is improved by including the per-photon energy resolution estimator $\sigma_E$ into the set of variables corrected with the normalizing flow (cf.~Section~\ref{sec:corr}).
Figure~\ref{fig:sigma_m_over_m_decorrelated} shows the agreement between the $\sigma_m / m$ distributions in \Zee data and simulation before and after propagating the corrections for $\sigma_E$ to $\sigma_m / m$.
Both electrons in the \Zee events are reconstructed as photons and the simulated events are reweighted to data in \pt, $\eta$, and $\phi$ of \gamProbee, as well as in $\rho$.
The corrected MC distribution agrees with the distribution observed in data within the assigned systematic uncertainty in $\sigma_E/E$ (5\%), as described in Section~\ref{sec:systematic_uncertainties}.

Three $\sigma_m/m$ categories are used in this analysis, as the improvement in sensitivity quickly saturates and does not noticeably improve with additional categories.
The resulting category boundaries are $[0,0.0105)$, $[0.0105,0.0130)$, and $[0.0130,\infty)$.
In all three categories, the minimum value for the photon ID BDT score is 0.25.

The combined acceptance and efficiency of the event selection described in this section is estimated from simulation to be $\approx$32\% for the ggH and VBF processes, about 27\% for the \VH process, and $\approx$28\% for the \ttH process with respect to the total phase space.

\section{Fiducial phase space and observables}
\label{sec:psDef_observables}
The cross section for the Higgs boson production is measured in a fiducial phase space defined at the particle level, with the goal of reducing model dependence and extrapolation uncertainties.
The fiducial criteria at the particle level are close to the event selection requirements at the detector level (Section~\ref{sec:selection_categorization}).
The criteria are based on the two \pt-leading photons in the event.
They must be within the fiducial acceptance of $\abs{\eta} < 2.5$ with $\ECALLow < \abs{\eta} < \ECALHigh$ excluded to match the rejection of the ECAL barrel-endcap transition regions at the detector level.
The photons must fulfil $\mathcal{I} < 10\GeV$, where the isolation variable $\mathcal{I}$ is the scalar \pt sum of stable, visible final-state particles in a cone of radius 0.3 centred on the photon momentum direction.
The geometric requirement of $\sqrt{\smash[b]{\ptGamLead \ptGamSublead}} / \mgg > 1/3$ is applied, and the subleading photon must fulfil $\ptGamSublead / \mgg > 1/4$.
The efficiency of these criteria, as determined from simulation, is $\approx$50.6\%.
This is slightly lower compared to the efficiency of $\approx51.8\%$ that is obtained with the requirement $\ptGamLead / \mgg >  1/3$ used in previous measurements~\cite{CMS:2018ctp,CMS:2022wpo} instead of the geometric requirement.

Fiducial cross sections are also measured in bins of four kinematic observables: the transverse momentum, the absolute value of the rapidity of the diphoton system, the number of jets (\NJ) in the event, and the transverse momentum of the leading jet (\PTJEins), defined as the jet with the highest \pt.
Jets at the fiducial level are built with the anti-\kt clustering algorithm out of stable particles with a distance parameter of 0.4, excluding neutrinos.
Jets are retained if they satisfy $\pt^\text{jet} > 30\GeV$, $\abs{\eta^\text{jet}} < 2.5$ and if they do not overlap with an electron or muon within $\DR < 0.4$, where the leptons have to fulfil the following requirements:
Electrons (muons) must have $\pt > 15\,(10)\GeV$, $\abs{\eta} < 2.5\,(2.4$), and $\mathcal{I}/\pt < 0.2$.
These criteria match the detector-level jet selections.
The bin boundaries are chosen to provide an expected relative uncertainty in each bin of about 40 (60\%) for \ptH and \absYH (\NJ and \PTJEins).
The boundary values are shown in Table~\ref{table:binning}.

\begin{table}[h!t]
    \topcaption{Bin boundaries for the differential cross section measurement. The first \PTJEins bin corresponds to events without jets. For the \NJ binning, the right boundary should be considered as not included in the bin, \ie [lower, upper).}
    \label{table:binning}
    \centering
    \begin{tabular}{l  {c}@{\hspace*{5pt}} c c c c c c c c c c}
        Observable && \multicolumn{10}{c}{Bin boundaries} \\
        \cline{1-1}\cline{3-12}
        \ptH (\GeVns{}) && 0 & 15 & 30 & 45 & 80 & 120 & 200 & 350 & $\infty$ & \rule{0pt}{2.5ex} \\
        \absYH && 0 & 0.15 & 0.3 & 0.6 & 0.9 & 2.5 & \multicolumn{4}{c}{} \\
        \NJ && 0 & 1 & 2 & 3 & $\infty$ & \multicolumn{5}{c}{} \\
        \PTJEins (\GeVns{}) && 0-jet & 30 & 75 & 120 & 200 & $\infty$ & \multicolumn{4}{c}{} \\
    \end{tabular}
\end{table}

\section{Statistical analysis}
\label{sec:statistical_analysis}
To extract the inclusive fiducial cross section, \sigFid, a binned profile likelihood fit to the \mgg distributions in the three mass-resolution categories is performed with the \textsc{Combine} tool~\cite{CMS:2024onh}.
A bin width of 0.25\GeV has been chosen, which is sufficiently small compared to the typical mass resolution of 1.5--2\GeV of the SM Higgs boson signal as indicated in Fig.~\ref{fig:signal_models}.
Systematic uncertainties are treated as nuisance parameters with Gaussian constraints.
The likelihood can be factorized over the $N_m=320$ bins of the \mgg distributions in each category:
\begin{multline}
    \label{eq:likelihood}
    \mathcal{L}(\sigFid, \vec{n}_{\text{bkg}} , \vec{\theta}_{\mathrm{S}}, \vec{\theta}_{\mathrm{B}})  =
    \mathrm{Pdf}(\vec{\theta}_{\mathrm{S}}) \mathrm{Pdf}(\vec{\theta}_{\mathrm{B}})
    \prod_{i=1}^{N_{\text{cat}}} \Pois\left( n^i_{\text{ev}} | n_{\text{sig}}^i + n_{\text{bkg}}^i \right) \\
    \times \prod_{l=1}^{N_m}
    \left(
    \frac{\sigFid
    K^i(\vec{\theta}_{\mathrm{S}}) S^i(\mgg^l|\vec{\theta}_{\mathrm{S}}) L
    + n_{\mathrm{OOA}}^i S_{\mathrm{OOA}}^i(\mgg^l|\vec{\theta}_{\mathrm{S}})
    + n_{\text{bkg}}^i B^i (\mgg^l|\vec{\theta}_{\mathrm{B}})
    }
    { n_{\text{sig}}^i + n_{\text{bkg}}^i }
    \right)^{n^{li}_{\text{ev}}},
\end{multline}
where
\begin{itemize}
\item $\mathrm{Pdf}(\vec{\theta})$ is the probability density function for the vector of nuisance parameters $\vec{\theta}$;
\item $\vec{\theta}_{\text{S}}$ and $\vec{\theta}_{\text{B}}$ are the vectors of nuisance parameters associated with the signal and the background models, respectively;
\item $\Pois(n|\lambda)$ is the probability mass function of the Poisson distribution for $n$ occurrences with an expectation value of $\lambda$;
\item $n_{\text{ev}}^i$ is the number of observed data events in category $i$ and $n_{\text{ev}}^{li}$ is the number of observed data events in category $i$ and \mgg\ bin $l$;
\item $n_{\text{sig}}^i (\sigFid, \vec{\theta}_{\text{S}}) = \sigFid K^i(\vec{\theta}_{\text{S}}) L + n_{\mathrm{OOA}}^i$ and $n_{\text{bkg}}^i$ are the number of signal and background events in category $i$, and $\vec{n}_{\text{bkg}}$ is the vector with entries $n_{\text{bkg}}^i$;
\item $N_{\text{cat}} = 3$ is the number of mass-resolution categories;
\item $K^i(\vec{\theta}_{\text{S}})$ is the efficiency for reconstructing an event in category~$i$;
\item $S^i(\mgg|\vec{\theta}_{\mathrm{S}})$ and $B^i(\mgg|\vec{\theta}_{\mathrm{B}})$ are the signal and background probability density functions in category $i$ (cf. Sections~\ref{sec:sigmodel} and~\ref{sec:bkgmodel});
\item $\mgg^l$ is the centre of the $l$-th \mgg\ bin;
\item $L$ is the integrated luminosity;
\item $n_{\mathrm{OOA}}^i S_{\mathrm{OOA}}^i(\mgg|\vec{\theta}_{\mathrm{S}})$ is the out-of-acceptance (OOA) signal contribution in category $i$, where $n_{\mathrm{OOA}}^i$ is the number of OOA events and $S_{\mathrm{OOA}}^i$ the corresponding \mgg signal model. $n_{\mathrm{OOA}}^i$ and $S_{\mathrm{OOA}}^i$ are estimated from simulation and are affected by the same set of nuisance parameters as in-fiducial signal events.
\end{itemize}

A similar approach is used to extract the differential cross sections.
In those fits, \sigFid is promoted to a vector of fiducial cross sections measured in the particle-level bins of a specific observable.
At the detector level, the same binning is used.
Consequently, all quantities carrying the index $i$ in Eq.~\eqref{eq:likelihood} are extended with an additional index $j$ to enumerate the same number of reconstruction-level bins.
Similarly, several quantities are extended with an index $k$ for the respective particle-level bin.
For example, the per-category efficiency $K^i(\vec{\theta}_{\text{S}})$ becomes the detector response matrix $K^{ij}_k(\vec{\theta}_{\text{S}})$ that relates the events from a particle-level bin $k$ with a detector-level bin $j$ and category $i$.
This permits to encode the unfolding directly in the likelihood.
The condition numbers for the response matrices are less than ten, so no regularization is performed. 
The condition number is defined as the absolute value of the ratio between the largest and smallest matrix eigenvalues.

\subsection{Signal model}
\label{sec:sigmodel}

\begin{figure}
  \centering
      \includegraphics[width=0.48\textwidth]{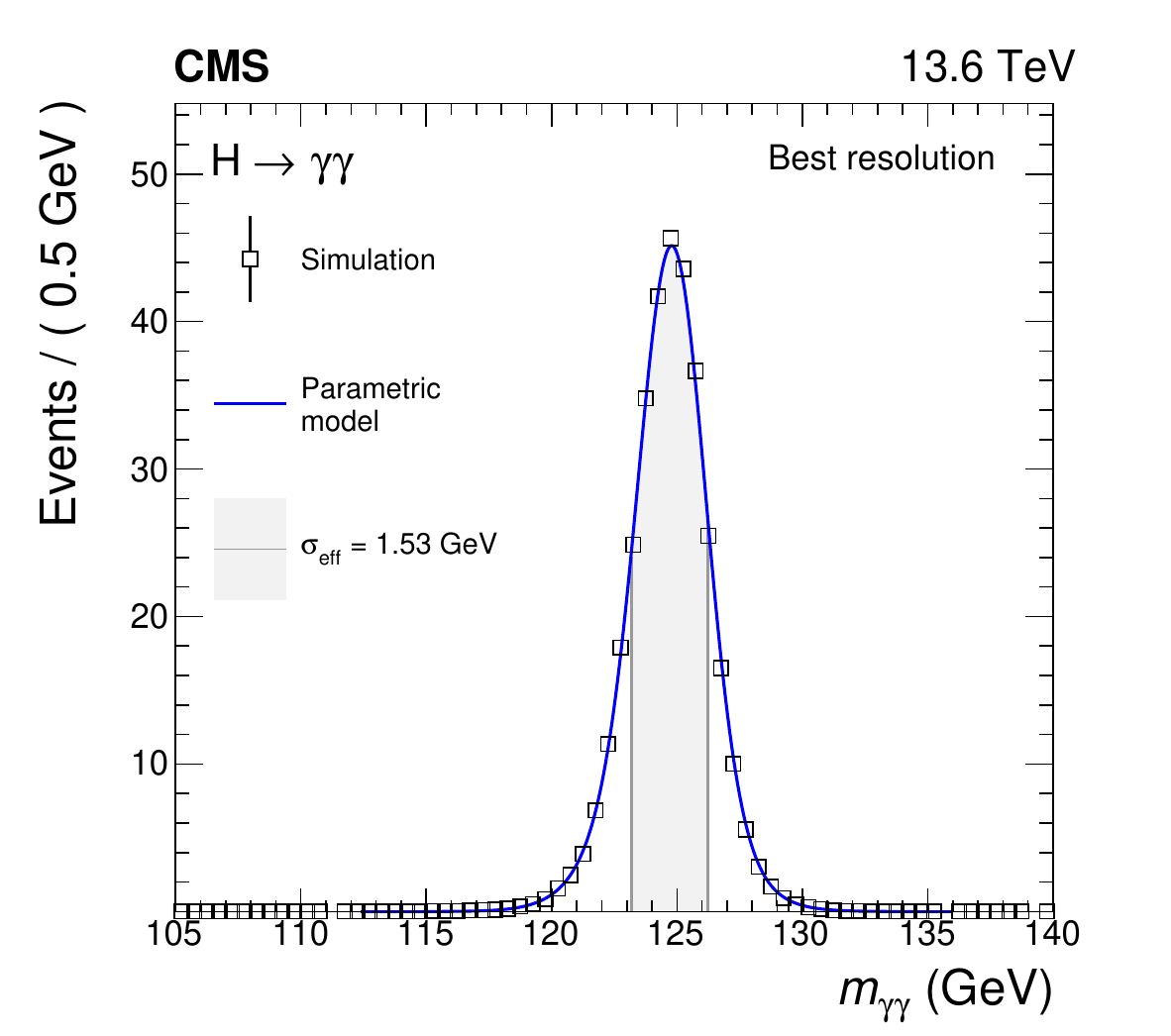}
      \includegraphics[width=0.48\textwidth]{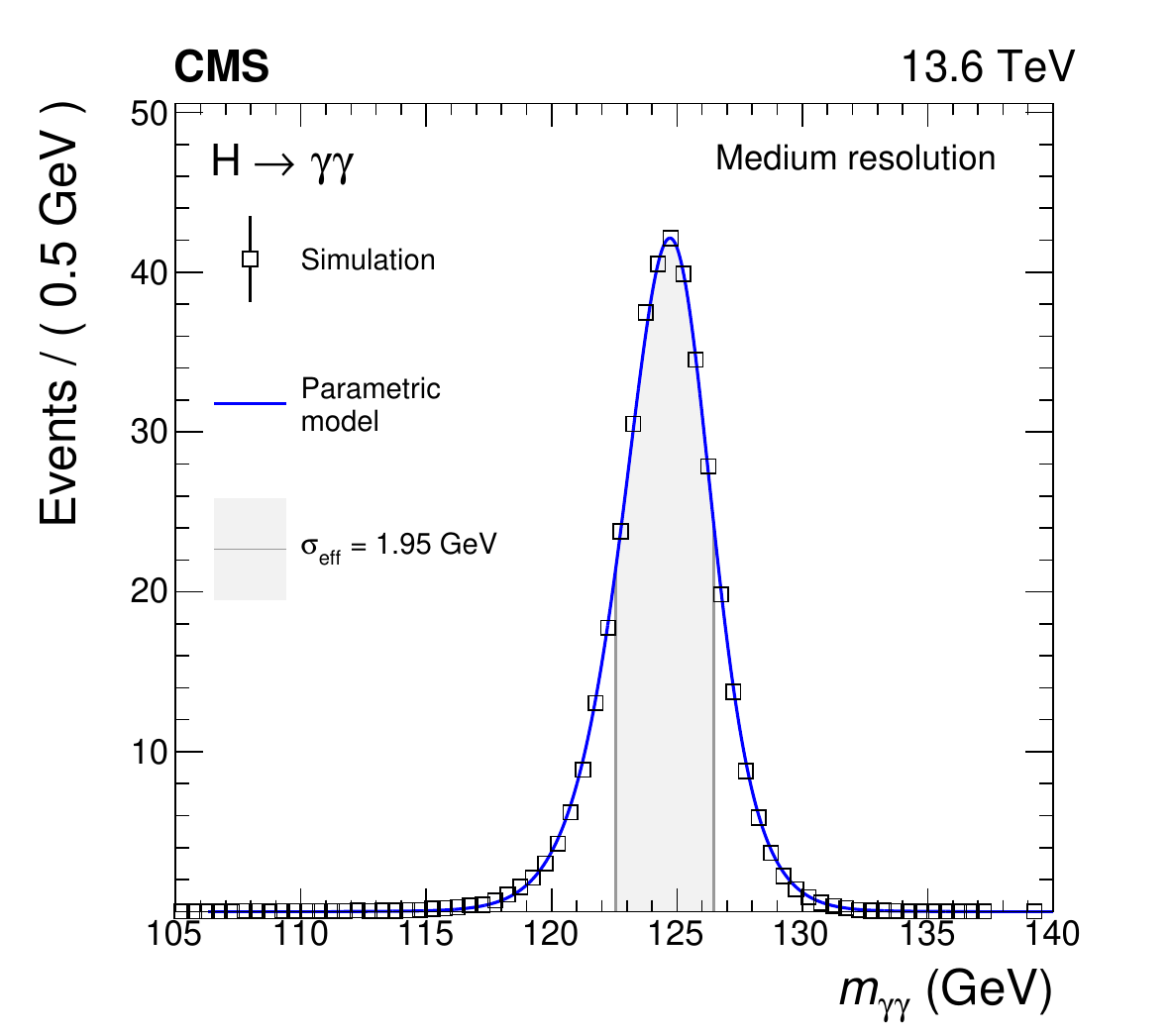}~\\
      \includegraphics[width=0.48\textwidth]{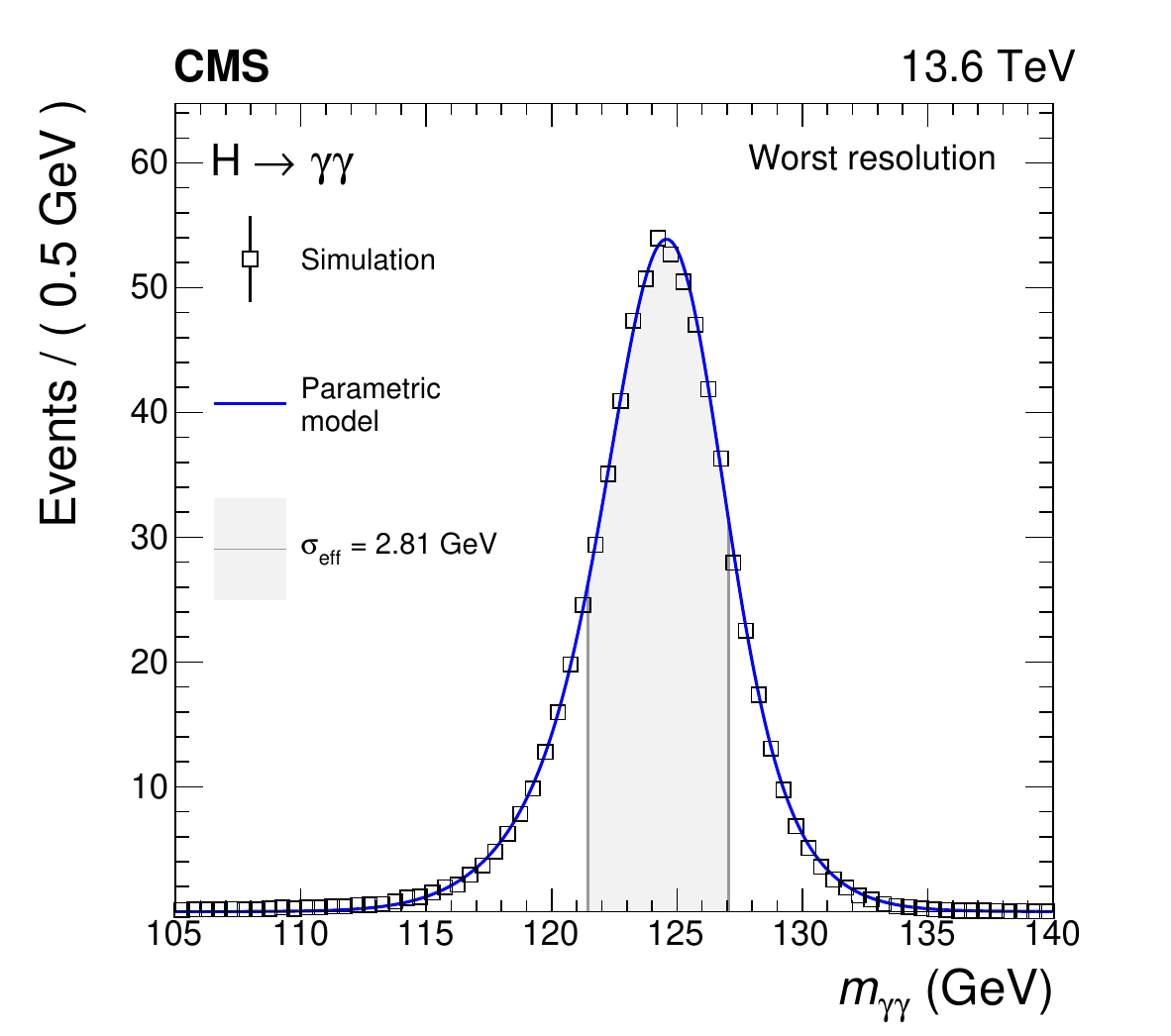}
      \includegraphics[width=0.48\textwidth]{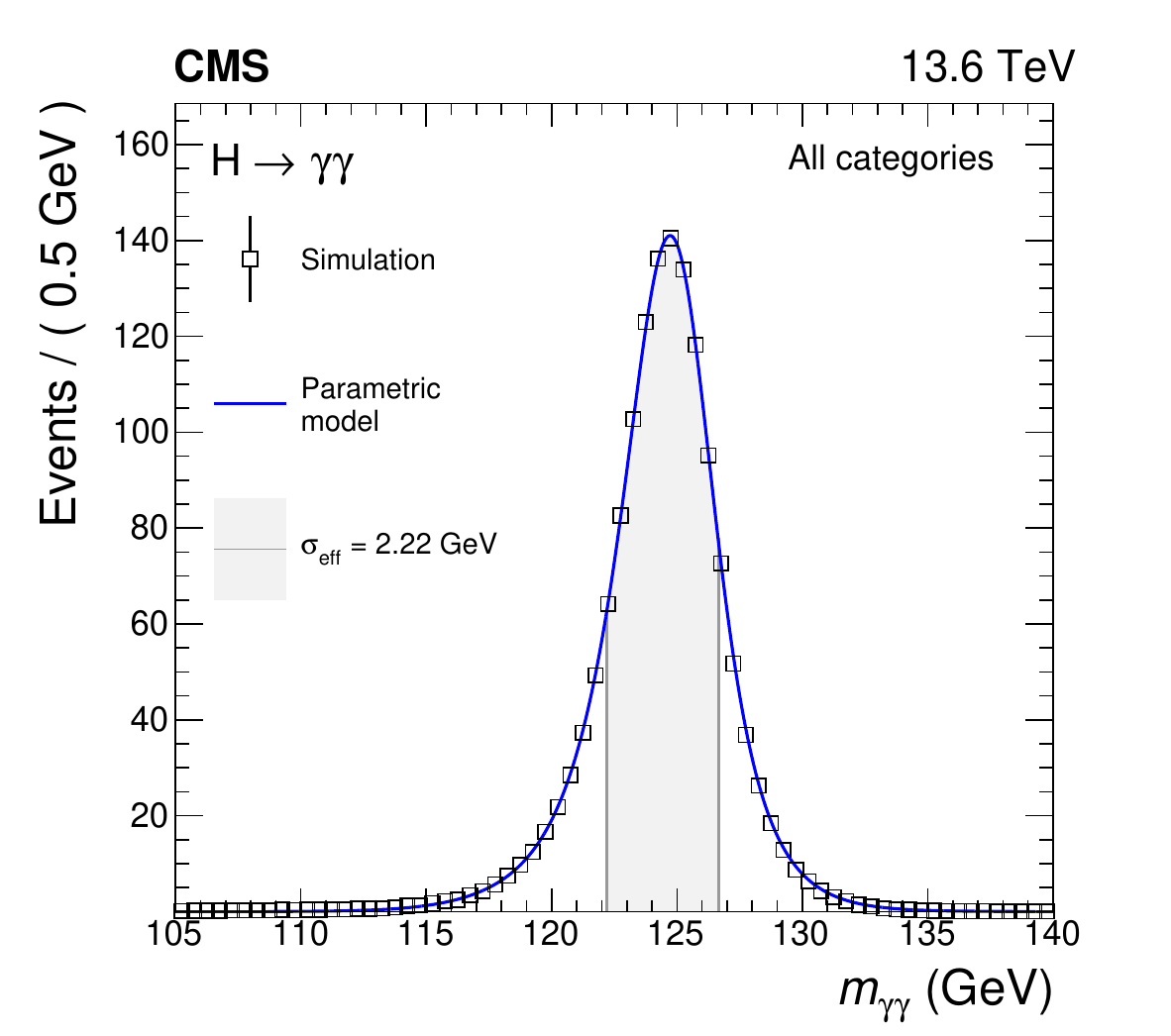}
  \caption{Combined parametrized signal shapes per category and for the sum of all categories for the measurement of the inclusive cross section.
  The open squares denote the expectation from the simulation and the blue lines show the parametric models that describe the simulations.
  The uncertainty bars for the expectation from the simulation due to the limited number of simulated events are smaller than the marker size.
  The normalization of the histograms corresponds to the expected number of events, taking into account the cross sections of the considered production modes, the efficiency of the selection, and the integrated luminosity of 34.7\fbinv.
  The effective mass resolution \sigmaEff (defined as half of the width of the smallest interval containing 68.3\% of the area of the distribution) for each combined signal model is indicated in the grey area.}
  \label{fig:signal_models}
\end{figure}

The signal model is based on simulated events, after including all the corrections described before.
A sum of up to five Gaussian functions is used as a parametric model to describe the \mgg resonant shape for each combination of reconstruction-level bins and $\sigma_m / m$ categories, separately for events passing the fiducial selection and the OOA contribution for each of the four main SM Higgs boson production modes.
The fraction of OOA events is the largest for the \VH and \ttH production processes, reaching up to 3.4\% in the worst-resolution category for the measurement of the inclusive cross section.
In the best-resolution category, the fraction of OOA events is below 1\% for the ggH, VBF, and \VH processes.
Additionally, different signal shapes are constructed for the cases of correct and incorrect identification of the primary vertex, as the invariant diphoton mass depends on the estimated photon direction and therefore the \mgg shape differs significantly for the cases of a correctly or incorrectly identified primary vertex.
The models for the correct and incorrect vertex identification are combined with their relative weights obtained from simulation.
This enables the inclusion of systematic uncertainties related to the vertex assignment in the fit (cf. Section~\ref{sec:systematic_uncertainties}).
To improve the modelling of the luminous region, its spread along the beam axis is reweighted in simulation to correspond to the spread in data.
The final nominal signal model is given by the weighted sum of the individual parametric models according to the respective SM cross section predictions, OOA fractions, and selection efficiencies.
The combined parametric signal models for the three mass-resolution categories and the weighted sum across all categories are shown in Fig.~\ref{fig:signal_models}.
The values of the effective mass resolution \sigmaEff, defined as half of the smallest interval centred around the mean that contains 68.3\% of the total area under the signal model histogram, are indicated.
Signal shapes are constructed for Higgs boson masses of 120, 125 and 130\GeV.
The parameters of the models for masses between these points are obtained using piecewise linear interpolation.
A piecewise cubic interpolation is used to determine the $\mathcal{B}(\Hgg)$, the acceptance and efficiency, and the right-vertex fraction as a function of the Higgs boson mass based on the same mass hypotheses.
The nominal Higgs boson mass is set to 125.38\GeV~\cite{CMS:2020xrn}.
The interference between the \Hgg signal and the continuous diphoton background~\cite{Campbell:2017rke} is not taken into account as it is at the percent level for an SM Higgs boson, which is well below the theoretical uncertainties.

\subsection{Background model}
\label{sec:bkgmodel}

The background model is defined in a data-driven way using the discrete profiling method~\cite{Dauncey:2014xga}.
Parametrized functions are used to model the smoothly falling background as a function of \mgg in the range 100--180\GeV.
A different model is constructed for each category.
During the extraction of the fiducial cross sections from the fit to data (cf.~Eq.~\eqref{eq:likelihood}), the choice of the background function is treated as a discrete nuisance parameter and accounts for the imperfect a-priori knowledge of the \mgg background shape.
The resulting uncertainty is absorbed into the statistical uncertainty.

Initial fits are performed over the whole \mgg range to determine the set of plausible function choices.
The considered functions are grouped into families: sums of Bernstein polynomials, sums of exponential functions, Laurent series, and sums of power-law functions.
In each family, multiple functions with different numbers of parameters are considered.
An $F$-test~\cite{fisher_chi2_1922} is performed to determine the maximum number of parameters to be used, whereas the minimum number is determined by a requirement on the goodness-of-fit to the data.
A higher number of degrees of freedom for the fitting function is penalized by constructing the likelihood using $-2\ln\mathcal{L}_B + N_B$, where $\mathcal{L}_B$ refers to the likelihood as a function of the background function parameters and $N_B$ is the number of free parameters of a given background function.
In the profile likelihood fit, at least four background functions are considered for each category with at least one function per family.
In the measurement of the inclusive cross section, the sum of three exponential functions, the Bernstein polynomial sum of fourth degree, and the sum of three exponential functions are the best-fit functions in the best-resolution, medium-resolution and worst-resolution categories, respectively.
For the measurement of the differential cross sections, a best-fit background function is determined for the three mass-resolution categories in each detector-level bin of the differential distribution.

\section{Systematic uncertainties}
\label{sec:systematic_uncertainties}
The following systematic uncertainties affect the signal model by allowing changes in the location and the width of the Higgs boson peak (detailed in Section~\ref{sec:sigmodel}):

\begin{itemize}
    \item \textit{Photon energy scale and resolution}: This is the uncertainty related to the scale of the photon energy in data and the resolution corrections in simulation.
    The total uncertainty consists of three components:
    First, a uniform $0.1\%$ uncertainty accounts for biases from the method related to the energy dependence of the resolution corrections to the simulation, efficiency scale factors, and kinematic differences between data and simulation.
    Additionally, the variations of the corrections from changing the fit window of the invariant dielectron mass from $[80, 100]$ to $[70, 110]\GeV$ are considered.
    Furthermore, the derivation of the corrections is repeated with a photon ID BDT score requirement of ${>}-0.9$ instead of ${>}0.25$ to estimate the impact of background contributions as well as the correlations between the energy and the photon ID BDT score.
    The energy scale uncertainties are the lowest for photons in the inner EB ($\abs{\eta} < 1$) and amount to $\approx0.1\%$ in this region.
    The relative uncertainty in the resolution term is around $5\%$ for high-\RNINE photons in the inner EB.
    It reaches $20$ ($40$)\% for high- (low-)\RNINE photons in the EE.
    Six uncorrelated nuisance parameters parametrize this source of uncertainty in the statistical model.
    \item \textit{\GEANTfour electromagnetic shower modelling}: An uncertainty is introduced to account for the imperfect modelling of electromagnetic showers in \GEANTfour.
    A simulation with an alternative shower description modifies the energy scale for both electrons and photons.
    This results in an uncertainty of $0.05\%$ in the photon energy scale.
    \item \textit{Non-uniformity of light collection}: This uncertainty is related to the modelling of the light collection depending on the emission depth in the ECAL crystals, which is different for electrons and photons~\cite{CMS:2020xrn}.
    It is only considered for photons with $\RNINE>0.975$ as most photons with smaller \RNINE values convert before reaching the calorimeter and therefore deposit most of their energy as electrons.
    This uncertainty affects the photon energy scale and amounts to about 0.14 (0.31)\% for photons in the EB (EE).
    \item \textit{Modelling of the material upstream of the ECAL}:
    The fraction of photons that convert before reaching the ECAL depends on the material before the ECAL whose description in the simulation is not perfect. 
    The impact on the photon energy scale is estimated by varying the amount of upstream material in the simulation.
    It amounts to 0.02--0.05\% for photons in the EB and reaches $0.25$\% for photons in the EE.
    \item \textit{Vertex assignment}: The fraction of events with a reconstructed vertex within $\lvert \Delta z \rvert < 1\,\mathrm{cm}$ of the true vertex, \ie with the mass resolution driven by the energy resolution of the photons, is varied by $\pm2\%$.
    The estimation of this uncertainty is detailed in Ref.~\cite{CMS:2021kom}.
\end{itemize}

Other sources of experimental systematic uncertainty affect the event yields. Their effect is parametrized with log-normal distributions. The following sources of experimental uncertainty are considered:

\begin{itemize}
    \item \textit{Integrated luminosity}: The uncertainty in the integrated luminosity collected in 2022 is $1.4\%$~\cite{LUM-22-001}.
    \item \textit{Cross section for pileup reweighting}: 
    Simulated events are reweighted so that the distribution of the number of reconstructed primary vertices in MC matches that observed in data.
    The data pileup distribution is inferred from the recorded instantaneous luminosity profile together with an assumed inelastic pp cross section of $69.2\,\mathrm{mb}$ and compared to that in simulation to determine the nominal reweighting factors.
    Its uncertainty of $3.2\,\mathrm{mb}$ is propagated by recomputing the data pileup distribution, yielding alternative event weights that are propagated to the statistical analysis and parameterized with a single nuisance parameter.
    This uncertainty primarily affects the extracted cross section through the correlation between pileup and the diphoton mass resolution estimator.
    \item \textit{Trigger efficiency}: The systematic uncertainty in the trigger efficiency is determined by using alternative background templates for the efficiency measurement from \Zee\ events. The uncertainty reaches $2.2\%$ for low-\RNINE\ photons, but it is generally below $0.3\%$ for photons in the EB and with high values of \RNINE, which is the case for most photons from Higgs boson decays that enter the best-resolution category.
    \item \textit{Photon preselection efficiency}: The uncertainty in the photon preselection scale factor is evaluated by varying the signal and background shapes used to determine the preselection efficiency in data and simulation from \Zee~ events and propagating these variations to the scale factors.
    The resulting uncertainty is generally below $1\%$ for photons in the EB with $\pt < 60\GeV$, where the \PZ\ boson decays provide a large number of events.
    For photons in the EE, the uncertainty can be as large as $2.8\%$.
    \item \textit{Electron veto efficiency}: \Zmmg events are used to determine the efficiency of the electron veto in both data and simulation.
    The limited number of events in data is the dominant uncertainty in the resulting scale factor.
    The uncertainty reaches $2.5\%$ for low-\RNINE\ photons in the EE, but is below $0.5\%$ otherwise.
    \item \textit{Photon identification efficiency}: After applying the normalizing-flow-based corrections, the photon ID BDT score distribution is in good agreement between data and simulation.
    Scale factors are calculated using the tag-and-probe method with \Zee\ events to correct the simulation for any remaining disagreement.
    The scale factors are compatible with unity within the uncertainties.
    Several sources of systematic uncertainty are taken into account, including alternative signal and background modelling templates.
    These uncertainties are combined and result in an uncertainty that is generally below $0.5\%$ for photons in the EB.
    For EE photons with low \RNINE, it can reach up to $2\%$.
    \item \textit{Per-photon energy resolution}: After applying the normalizing-flow-based correction, a small residual disagreement in the distribution of $\sigma_E/E$ between data and simulation remains for both \Zee\ events as well as diphoton events with $m_{\gamma \gamma} / \GeV \in [100,120] \cup [130,180]$.
    A conservative uncertainty of $5\%$ is applied to $\sigma_E/E$ and propagated to the invariant mass resolution of the diphoton system.
    This uncertainty mainly results in migrations between the mass-resolution categories.
    \item \textit{Jet energy correction and jet energy resolution}: The uncertainty in the calibration of the JEC and JER~\cite{CMS:2016lmd} directly affects the selection efficiency of the jets for the differential measurement of the number of jets.
    The uncertainty in the JEC and JER is 2--5\%, depending on the \pt and $\eta$ of the jets.
    \end{itemize}

In addition, several sources of theoretical uncertainty are taken into account.
The normalization of each particle-level bin is fixed while evaluating the effect of all of these uncertainties except for those related to the limited number of events in the simulation samples.
Thus, their effect only enters in category migrations between detector-level bins and mass-resolution categories.
The following sources are considered:
\begin{itemize}
    \item \textit{Parton distribution function uncertainties}: The uncertainty in the imperfect knowledge of the PDFs is estimated by reweighting events according to the NNPDF3.1~\cite{NNPDF:2017mvq} prescription using the compressed Hessian set of PDF eigenvectors.
    This source of uncertainty is parameterized with $100$ independent nuisance parameters in the statistical model.
    \item \textit{Renormalization and factorization scales uncertainty}: This uncertainty is related to the missing higher-order terms in the perturbation series for the cross section calculation.
    Simulated events are reweighted with alternative event weights where the scales are varied by a factor of two, excluding the $(2,1/2)$ and $(1/2,2)$ variations.
    \item \textit{\alpS uncertainty}: As the PDF uncertainty, the uncertainty in the value of the strong coupling constant is taken from the NNPDF3.1 prescription.
    It is evaluated by varying \alpS by $\pm 0.002$ from the nominal value of $\alpS (m_\PZ) = 0.118$.
    \item \textit{Parton shower uncertainty}: The uncertainty in the modelling of the parton shower is estimated with reweighting factors that correspond to per-event cross section variations with the scales for initial- and final-state radiation varied up and down by a factor of two.
\end{itemize}

Finally, the uncertainty due to the limited number of events in the simulation samples is accounted for and parametrized using one nuisance parameter per category and for each of the two periods before and after the ECAL cooling issue~\cite{Barlow:1993dm,Conway:2011in}.

The dominant sources of systematic uncertainties for the measurement of the fiducial cross section are summarized in Table~\ref{tab:systImpacts}.

\section{Results}
\label{sec:results}
\begin{figure}[htb]
    \centering
        \includegraphics[width=0.55\textwidth]{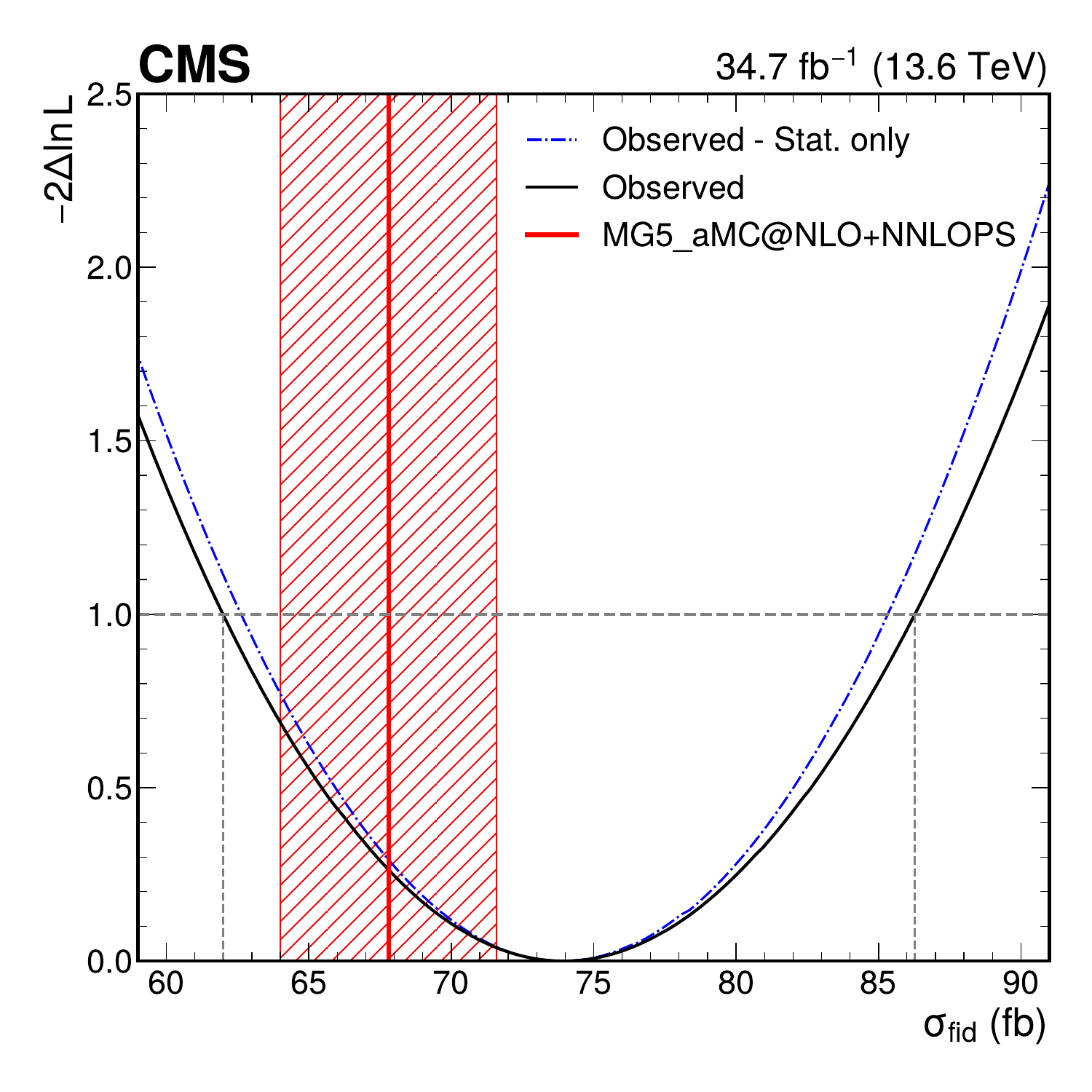}
        \caption{
        Likelihood scans for the inclusive fiducial cross section measurement.
        The black line corresponds to considering both the statistical and systematic uncertainties.
        The blue dash-dotted line corresponds to considering only the statistical uncertainty, including the discrete profiling method for the background modelling uncertainty.
        The theoretical prediction from \MGvATNLO, including the \NNLOPS reweighting for the \ggH component, is shown in red.
        The shaded theory uncertainty band includes the uncertainties in the renormalization and factorization scales, in the parton distribution functions, in \alpS, in the $\mathcal{B}(\Hgg)$, and in the fiducial acceptance.}
        \label{fig:xsec_likelihood_scan}
\end{figure}

The measured cross section in the fiducial phase space, defined in Section~\ref{sec:psDef_observables}, is 
\begin{equation}
    \sigFid = 74\pm11\stat^{+5}_{-4} \syst\fb= 74\pm12\fb.
\end{equation}
Figure~\ref{fig:xsec_likelihood_scan} shows the likelihood scans for the inclusive fiducial cross section measurement.
It also shows the theoretical prediction of $67.8 \pm 3.8 \fb = 67.8 \pm 2.6\,(\text{scales}) \pm 2.3\,(\mathrm{PDF}+\alpS) \pm 1.4\,(\mathcal{B})\fb$, calculated with \MGvATNLO reweighted to match the \NNLOPS prediction for ggH, and interfaced with {\PYTHIA}8 (version 8.240)~\cite{Sjostrand:2014zea} using the CP5 tune.
The measured value agrees with this prediction within the uncertainties. 
The uncertainty in the prediction combines contributions from the Higgs boson production cross sections, the $\mathcal{B}(\Hgg)$, and the fiducial acceptance.
The first two uncertainties are taken from Refs.~\cite{LHCHiggsCrossSectionWorkingGroup:2016ypw} and ~\cite{13p6_LHCHWG_extrap}, respectively, whereas the third is computed with the MC samples presented in Section~\ref{sec:samples}.
The variations of the fiducial acceptance are evaluated with the compressed Hessian eigenvector set of NNPDF3.1~\cite{NNPDF:2017mvq}, the \alpS\ value varied by 0.002 around its nominal value of 0.118, and the renormalization and factorization scales varied by a factor of 2, while excluding the $(2,1/2)$ and $(1/2,2)$ variations.

The dominant sources of systematic uncertainty for the measurement of the inclusive fiducial cross section are shown in Table~\ref{tab:systImpacts}.
The main contribution is from the per-photon energy resolution, related to migrations between the mass-resolution categories, followed by the photon energy scale and resolution, which affect the location and the width of the Higgs boson peak.
Other experimental uncertainties not listed in Table~\ref{tab:systImpacts} are below 0.5\% and the theoretical uncertainty is below 1\%.

\begin{table}[htb]
    \centering
    \topcaption{Magnitude of the systematic uncertainties (Impact) in the inclusive fiducial cross section measurement.
    The magnitude of the uncertainty from the photon energy scale and resolution is extracted by performing a fit with the corresponding group of nuisance parameters frozen to their best-fit values. The obtained confidence interval is then subtracted in quadrature from the total confidence interval from the fit where all nuisance parameters are profiled.
    The magnitudes of the other sources of systematic uncertainty are obtained by varying the corresponding nuisance parameter by one standard deviation, keeping the other nuisance parameters at their best-fit values.}
    \begin{tabular}{lc}
        Systematic uncertainty & Impact in \%\\\hline
        {Category migration from energy resolution} & $+3.5/{-4.2}$\\
        {Photon energy scale and resolution group} & $+3.4/{-2.8}$\\
        {Integrated luminosity} & $\pm1.4$\\
        {Photon preselection efficiency} & $\pm 1.4$\\
        {Material budget} & $+1.3/{-1.2}$\\
        {Photon identification efficiency} & $\pm 1.0$\\
        {Pileup reweighting} & $\pm$0.8\\
    \end{tabular}
    \label{tab:systImpacts}
\end{table}

\begin{figure}[htbp]
    \centering
        \includegraphics[width=0.55\textwidth]{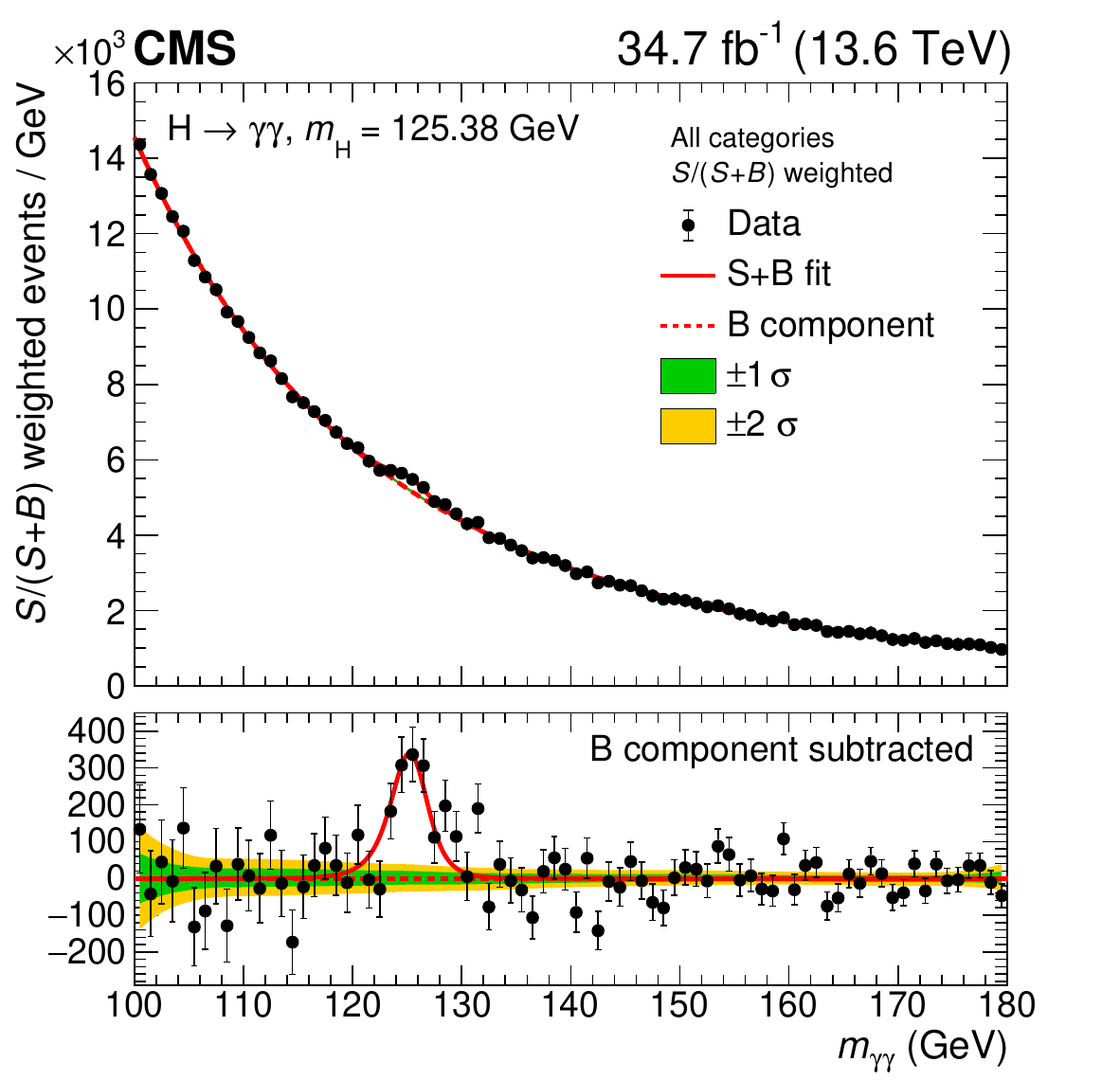}
        \caption{
        Diphoton invariant mass distribution in the inclusive fiducial measurement, weighted by $S/(S+B)$ for the different mass-resolution categories.
        The distribution is shown together with the signal$+$background fit (red line) and the background-only component (dashed line).
        In the lower panel, the signal component is shown, estimated by subtracting the background component from the signal$+$background fit.
        The green (yellow) bands indicate the $\pm1\sigma$ ($\pm2\sigma$) uncertainties in the background component.
        They are derived from pseudoexperiments using the best-fit background function from the signal+background fit.}
        \label{fig:mgg_fid_incl}
\end{figure}

The observed diphoton invariant mass distribution is shown in Fig.~\ref{fig:mgg_fid_incl} together with the combined signal and background fit and the background component alone.
The distributions from the three mass-resolution categories in the inclusive fiducial measurement are included and weighted by $S/(S+B)$, where $S$ and $B$ are the number of signal and background events in the central interval of width $2\sigmaEff$ around the fitted peak position per mass-resolution category.

\begin{figure}
    \centering
        \includegraphics[width=0.48\textwidth]{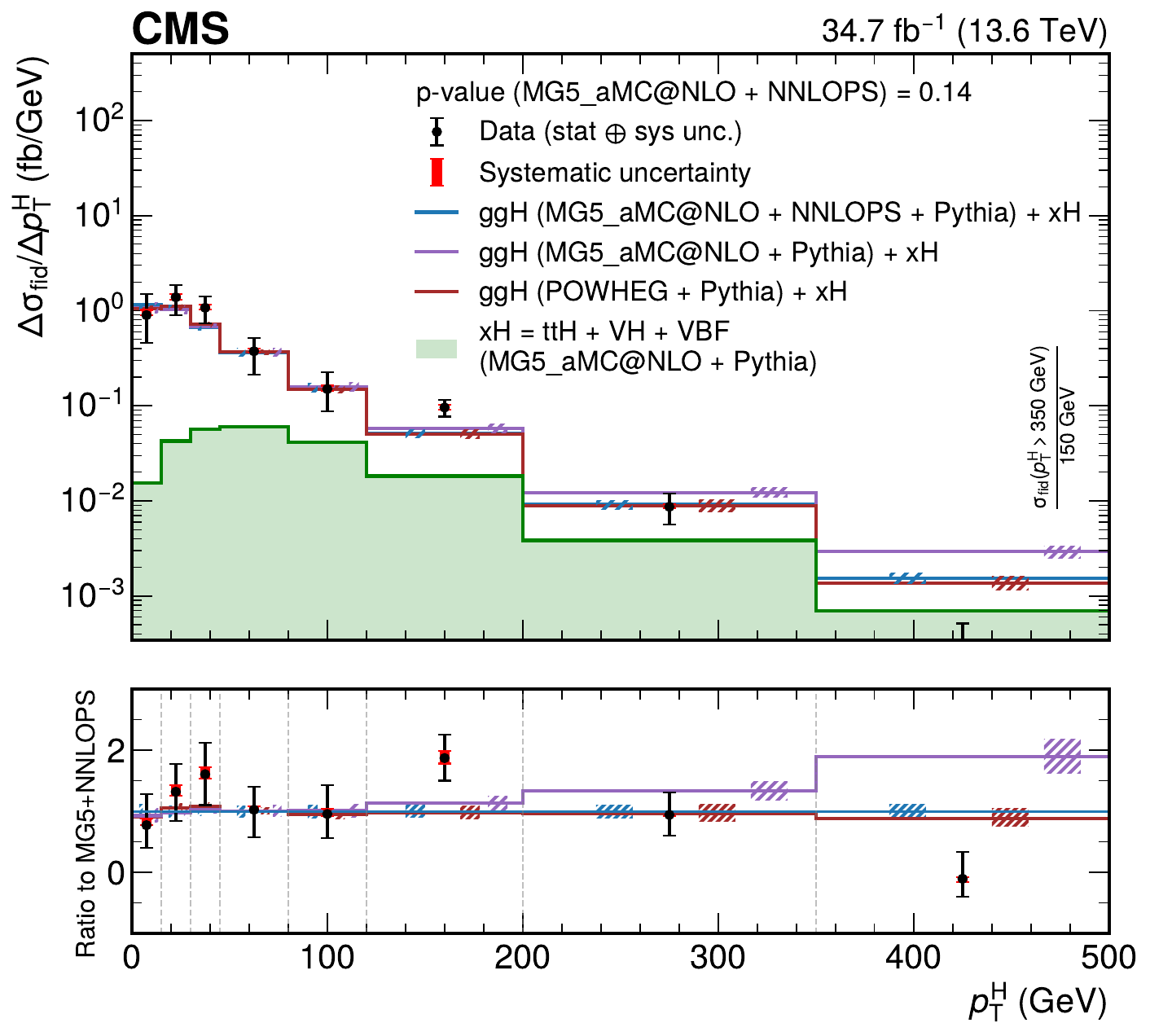}
        \includegraphics[width=0.48\textwidth]{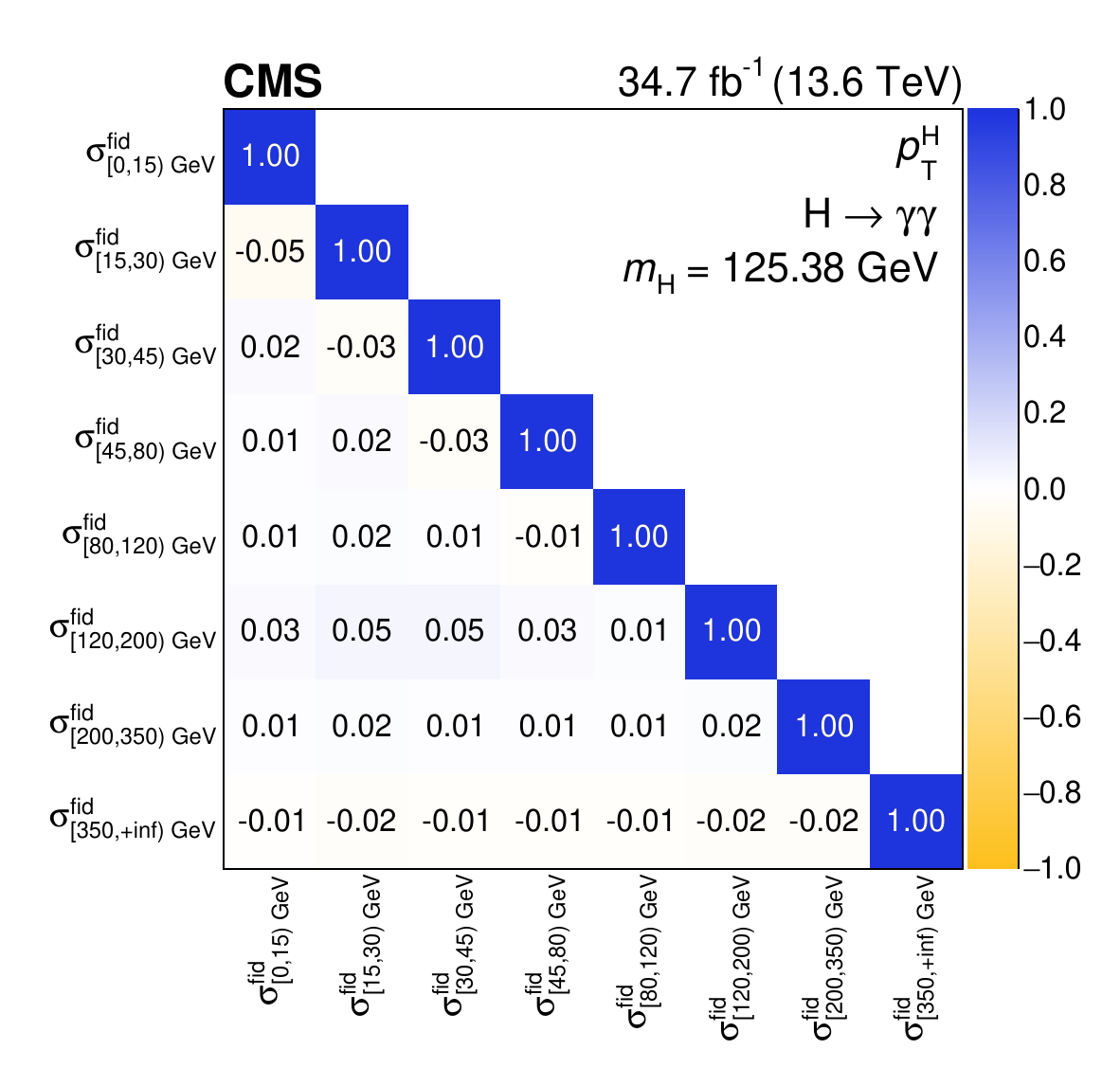}
        \caption{
            Differential fiducial cross sections for \ptH~(left) and the corresponding correlation matrix~(right).
            The measured cross section in each bin is divided by the corresponding bin width.
            The coloured lines denote the predictions from different event generation setups, explained in the legend and in the text.
            The dashed boxes show the uncertainties in theoretical predictions on both the \ggH and \xH components.
            The $p$-value is calculated for the nominal SM prediction, which is \MGvATNLO with \NNLOPS (MG5\_aMC@NLO + NNLOPS) reweighting.
            The lower panel in the left plot shows the ratio to the nominal SM prediction.
            The last bin extends to infinity and the normalization of the bin is indicated in the plot.
            }
        \label{fig:mgg_fid_pT}
\end{figure}

\begin{figure}
    \centering
        \includegraphics[width=0.48\textwidth]{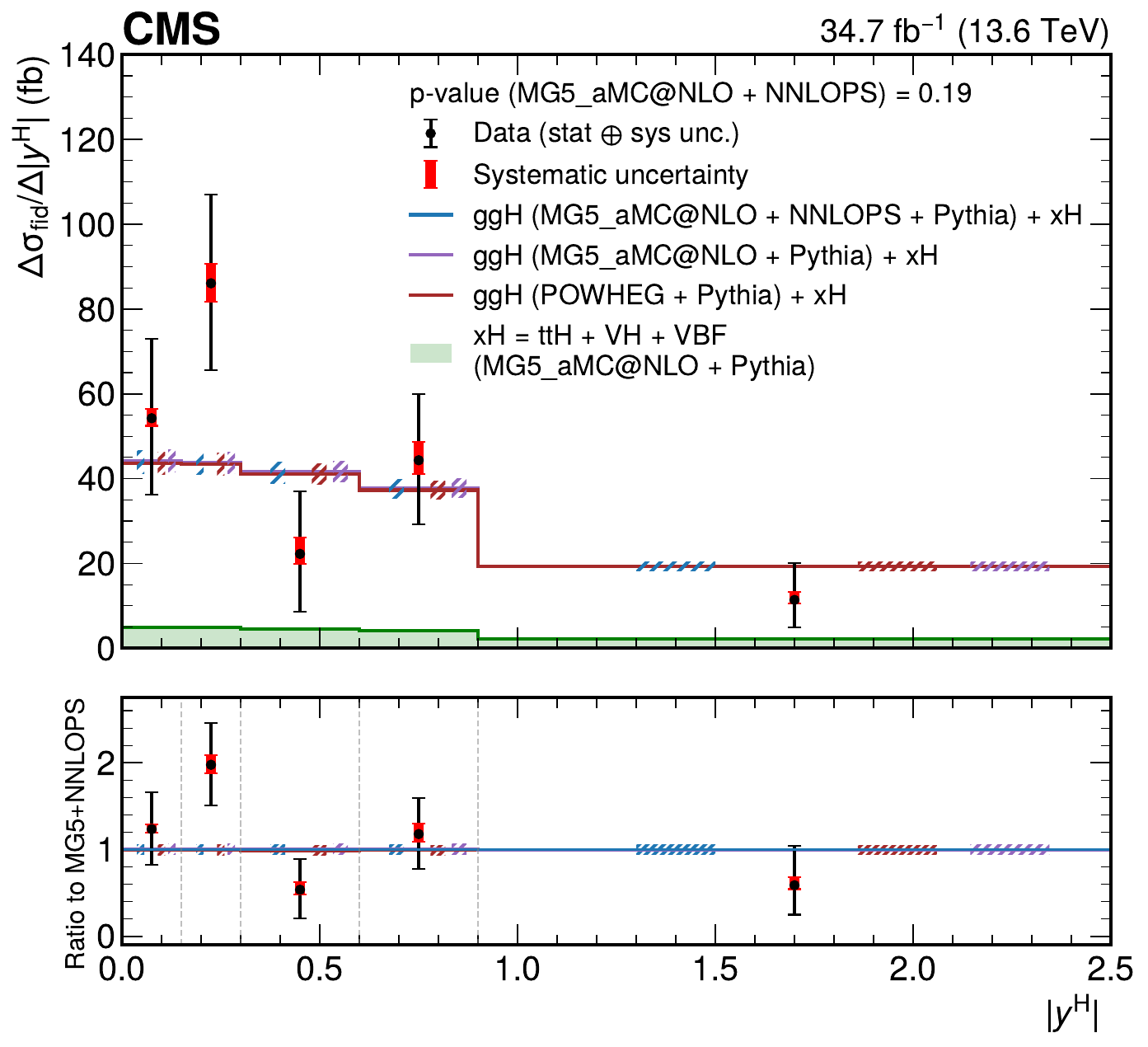}
        \includegraphics[width=0.48\textwidth]{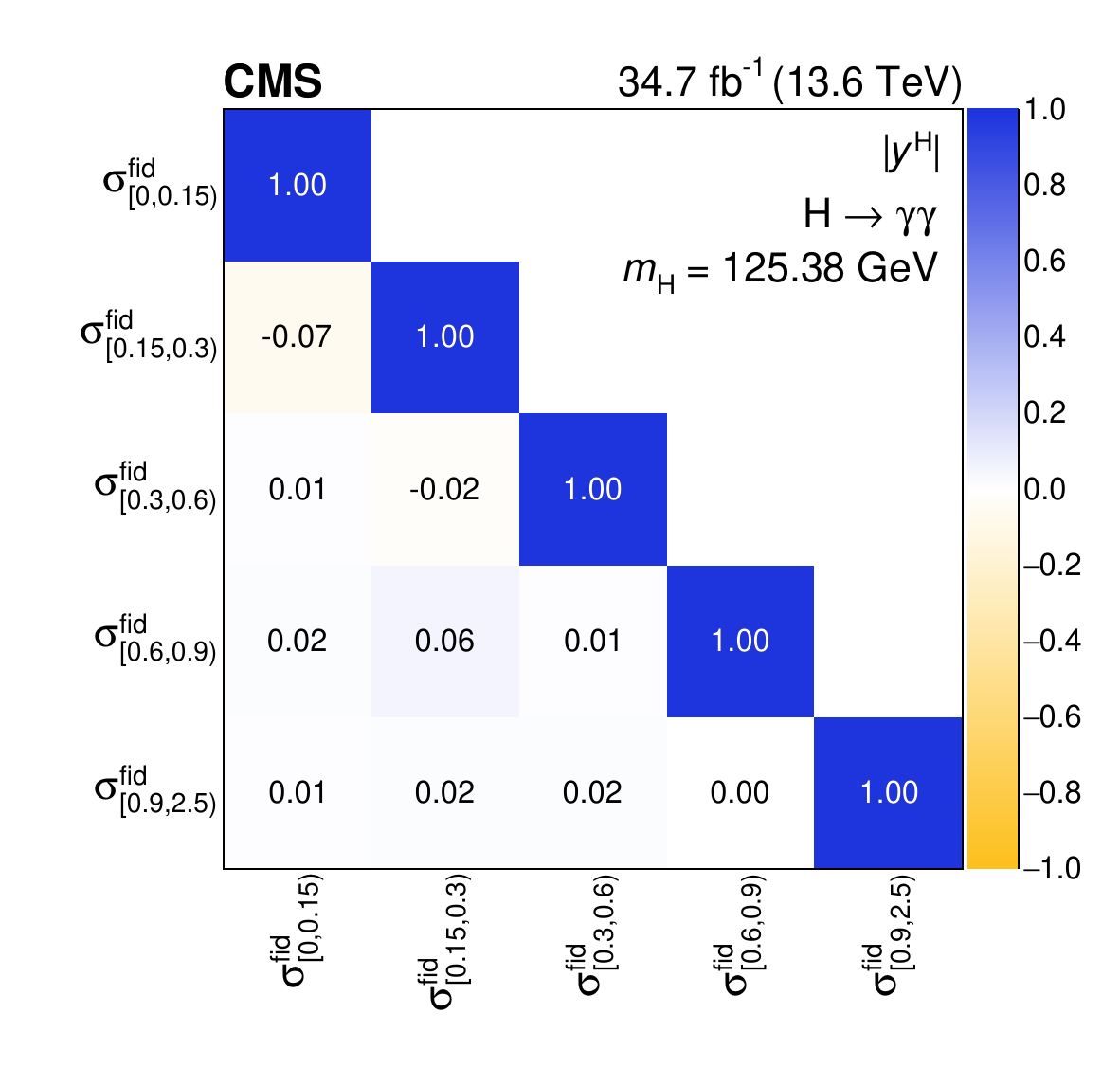}
        \caption{
            Differential fiducial cross sections for \absYH~(left) and the corresponding correlation matrix~(right).
            Other details as for the caption of Fig.~\ref{fig:mgg_fid_pT}. In this case, the last bin does not extend to infinity, but it is limited to 2.5.
        }
        \label{fig:mgg_fid_yh}
\end{figure}

\begin{figure}
    \centering
        \includegraphics[width=0.48\textwidth]{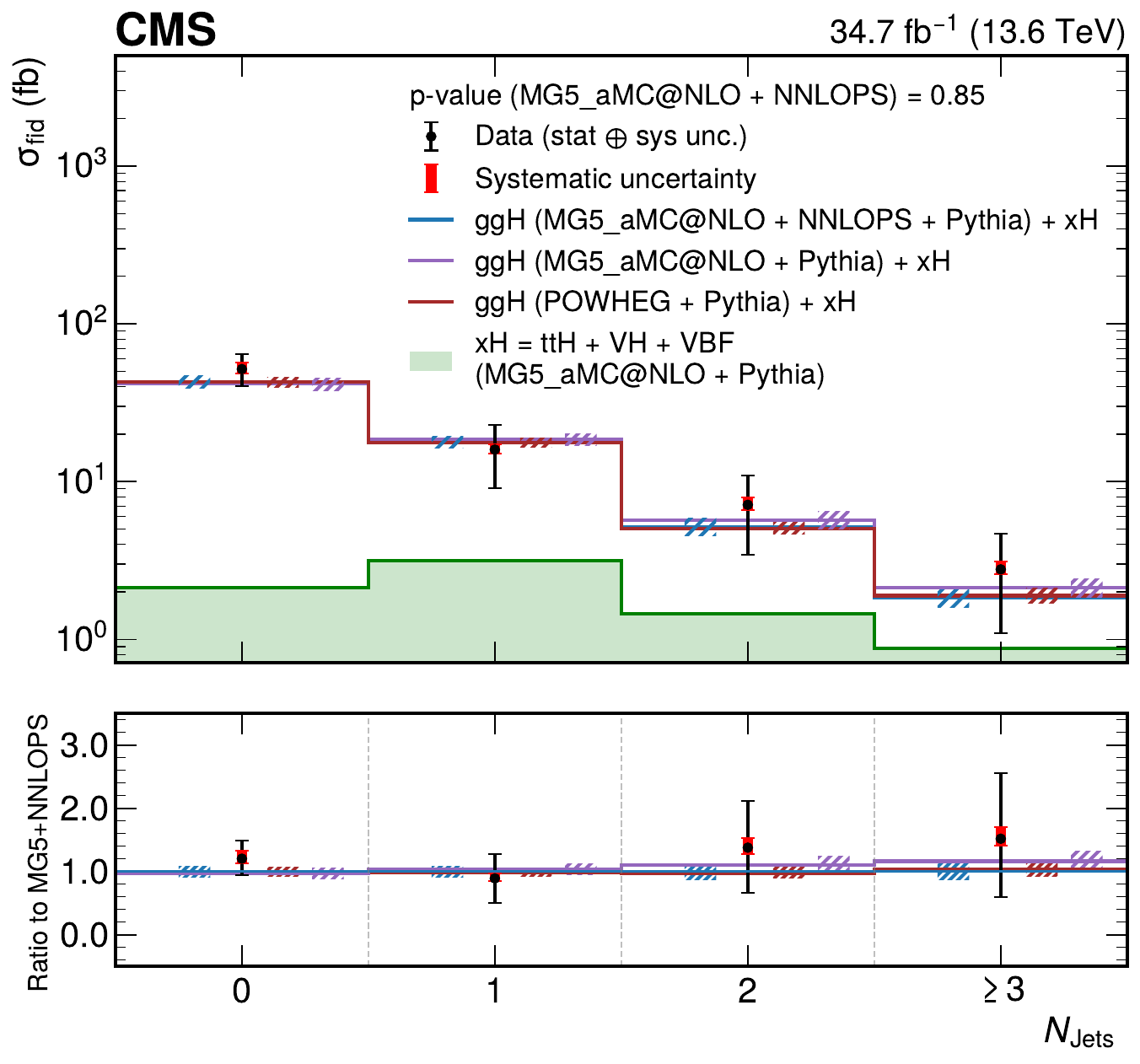}
        \includegraphics[width=0.48\textwidth]{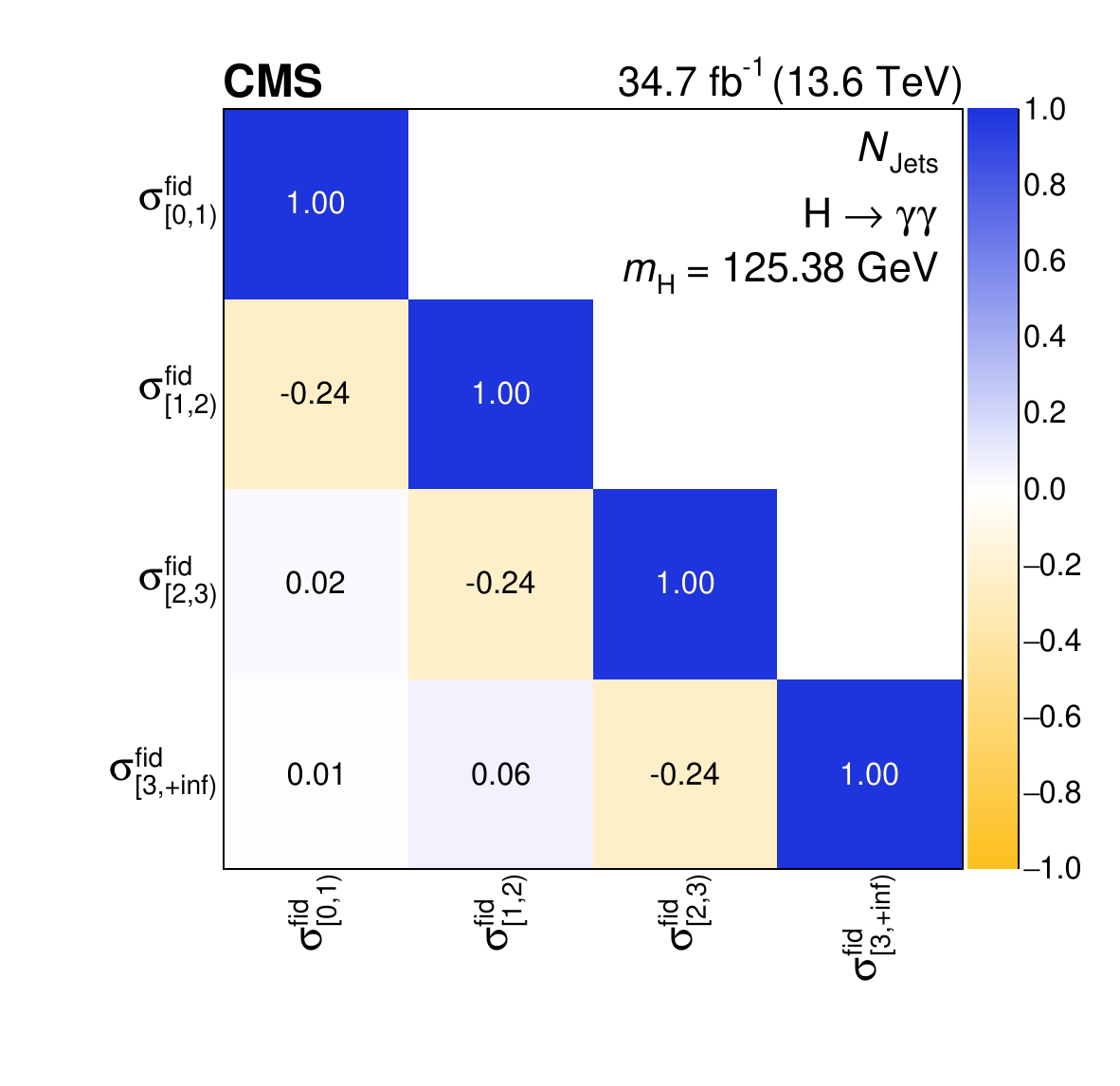}
        \caption{
            Differential fiducial cross sections for \NJ~(left) and the corresponding correlation matrix~(right).
            Other details as for the caption of Fig.~\ref{fig:mgg_fid_pT}.
        }
        \label{fig:mgg_fid_nj}
\end{figure}

\begin{figure}
    \centering
        \includegraphics[width=0.48\textwidth]{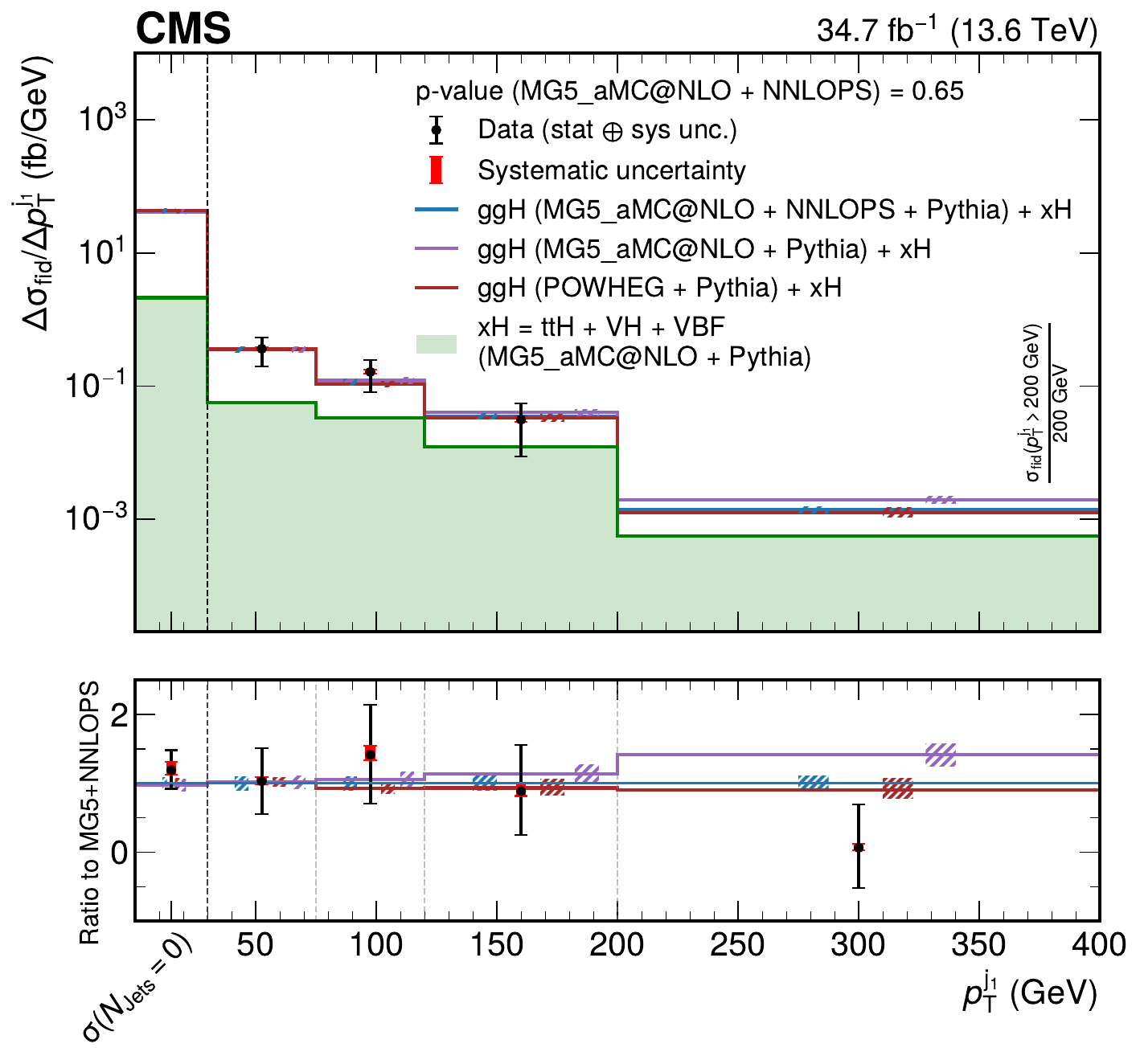}
        \includegraphics[width=0.48\textwidth]{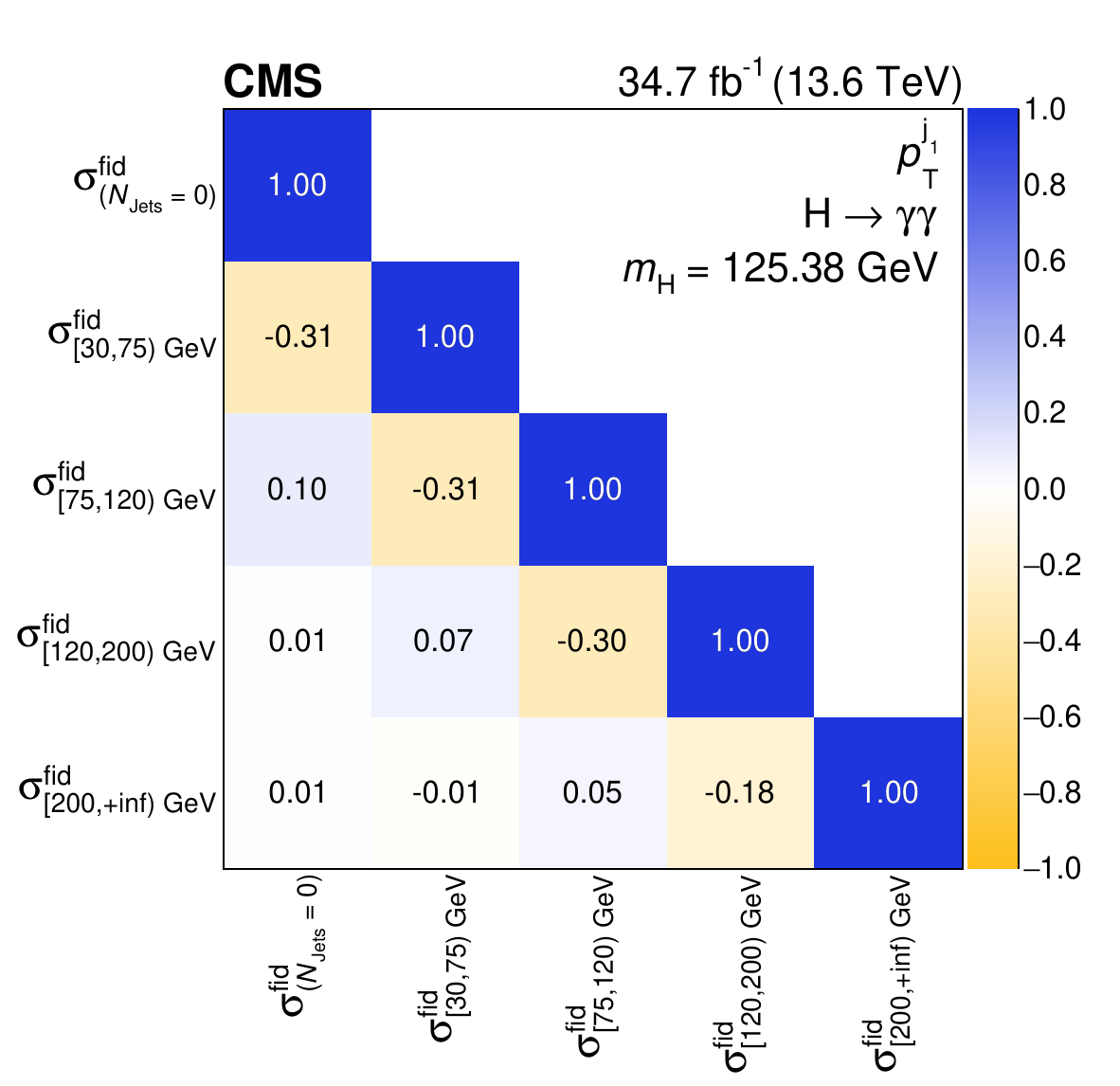}
        \caption{
            Differential fiducial cross sections for \PTJEins~(left) and the corresponding correlation matrix~(right).
            Other details as for the caption of Fig.~\ref{fig:mgg_fid_pT}.
        }
        \label{fig:mgg_fid_ptj1}
\end{figure}

The differential fiducial cross sections for the observables and the binning introduced in Section~\ref{sec:psDef_observables} are extracted from the maximum likelihood fit together with their uncertainties and the correlation matrices.
The spectra and correlation matrices for \ptH, \absYH, \NJ, and \PTJEins are presented in Figs.~\ref{fig:mgg_fid_pT},~\ref{fig:mgg_fid_yh}, \ref{fig:mgg_fid_nj}, and \ref{fig:mgg_fid_ptj1}, respectively. 
The cross sections measured in bins of jet-related observables exhibit stronger correlations due to the larger bin-by-bin migrations induced by the relatively poor energy resolution for jets compared to photons, whereas there is almost no correlation for the well measured variables \ptH and \absYH.
The measured differential cross sections are compared to various theoretical predictions.
For these theoretical predictions, the acceptances in the differential bins are calculated using the \ggH predictions from three different generators, whereas the inclusive prediction is normalized to the next-to-next-to-next-to-leading order computation \cite{LHCHiggsCrossSectionWorkingGroup:2016ypw,13p6_LHCHWG_extrap}.
The three predictions are taken from the \MGvATNLO simulation, with and without \NNLOPS reweighting, and from the \POWHEG 2.0 event generator~\cite{Nason:2004rx,Frixione:2007vw,Alioli:2010xd,Bagnaschi:2011tu}.
The acceptance of the $\xH=\VBF+\VH+\ttH$ component of the signal is taken from the \MGvATNLO simulation, whereas the inclusive prediction is normalized to the integrated cross sections reported in Ref.~\cite{13p6_LHCHWG_extrap}.
The uncertainties in the theoretical predictions are computed following the same strategy as for the inclusive cross section, described above.

To assess the compatibility with the SM predictions, $p$-values are computed for every observable from 
\begin{equation}
    p = \int^{\infty}_{\Delta} f(x; N_{\text{Bins}})\,\rd x,
    \label{eqn:pvalue}
\end{equation}
where $f(x; N_{\text{Bins}})$ is the probability density function for a chi-squared variable ($\chi^{2}$) with $N_{\text{Bins}}$ degrees of freedom and $\Delta = 2(\mathrm{NLL}(\vec{\sigma}_{\mathrm{SM}}) - \mathrm{NLL}(\vec{\hat{\sigma}})) = \chi^{2}_{\mathrm{SM}}$ is twice the difference between the negative log likelihood (NLL) evaluated for the SM hypothesis and for the best-fit values.
The $p$-values for the nominal prediction, \MGvATNLO\ with \NNLOPS reweighting, are 0.14, 0.19, 0.85, and 0.65 for the measurements of the \ptH, \absYH, \NJ, and \PTJEins differential cross sections, respectively.
These values are above 5\% and show good compatibility with the SM prediction.

The residual model dependence of the differential measurements is also tested.
Differential fiducial cross sections are extracted from fits to an Asimov data set~\cite{Cowan:2010js} that comprises the background component generated from the best-fit background modelling function and a signal component from the SM signal model.
In these fits, the signal component is assumed to be entirely composed of \VBF, \VH, or \ttH.
This introduces a model dependence, as the migration matrices do not correspond to the SM scenario and the normalization of the OOA component changes.
The differences with the extracted differential cross sections using the SM signal component hypothesis are below 12\% in every particle-level \ptH bin.
The average deviation is 8.3\%, which is much smaller than the expected statistical uncertainty per bin.
Thus, the presented differential cross section measurements are model independent within the statistical uncertainties given that the tested scenarios of 100\% \VBF, \VH, and \ttH contributions in the signal model are extreme considering the current experimental knowledge of Higgs boson production.

\section{Summary}
\label{sec:summary}
The fiducial inclusive cross section for Higgs boson production in proton-proton collisions has been measured at a centre-of-mass energy of 13.6\TeV using the \Hgg decay channel.
The data were collected with the CMS detector at the LHC and correspond to an integrated luminosity of 34.7\fbinv.
A new normalizing-flow-based method is applied to correct the imperfect modelling of reconstructed photon variables in the simulation and to reduce the associated systematic uncertainties.
The fiducial phase space is defined at the particle level and requires two isolated photons within the pseudorapidity $\abs{\eta}<2.5$ and not within $\ECALLow < \abs{\eta} < \ECALHigh$.
These photons must fulfil a requirement on the geometric mean of their transverse momenta scaled by their invariant mass, $\sqrt{\smash[b]{\ptGamLead \ptGamSublead}} / \mgg > 1/3$, which improves the perturbative convergence of the theoretical predictions, as well as the requirement $\ptGamSublead / \mgg > 1/4$.
The measured inclusive fiducial cross section is $\sigFid = 74\pm11\stat^{+5}_{-4}\syst\fb$ and is in agreement with the standard-model (SM) expectation of $67.8 \pm 3.8 \fb$.
Differential cross sections are measured as functions of the Higgs boson transverse momentum, rapidity, the number of associated jets, and the transverse momentum of the leading jet in the event.
Within the uncertainties, the differential cross sections agree with the SM predictions.

\begin{acknowledgments}
We congratulate our colleagues in the CERN accelerator departments for the excellent performance of the LHC and thank the technical and administrative staffs at CERN and at other CMS institutes for their contributions to the success of the CMS effort. In addition, we gratefully acknowledge the computing centres and personnel of the Worldwide LHC Computing Grid and other centres for delivering so effectively the computing infrastructure essential to our analyses. Finally, we acknowledge the enduring support for the construction and operation of the LHC, the CMS detector, and the supporting computing infrastructure provided by the following funding agencies: SC (Armenia), BMBWF and FWF (Austria); FNRS and FWO (Belgium); CNPq, CAPES, FAPERJ, FAPERGS, and FAPESP (Brazil); MES and BNSF (Bulgaria); CERN; CAS, MoST, and NSFC (China); MINCIENCIAS (Colombia); MSES and CSF (Croatia); RIF (Cyprus); SENESCYT (Ecuador); ERC PRG, RVTT3 and MoER TK202 (Estonia); Academy of Finland, MEC, and HIP (Finland); CEA and CNRS/IN2P3 (France); SRNSF (Georgia); BMBF, DFG, and HGF (Germany); GSRI (Greece); NKFIH (Hungary); DAE and DST (India); IPM (Iran); SFI (Ireland); INFN (Italy); MSIP and NRF (Republic of Korea); MES (Latvia); LMTLT (Lithuania); MOE and UM (Malaysia); BUAP, CINVESTAV, CONACYT, LNS, SEP, and UASLP-FAI (Mexico); MOS (Montenegro); MBIE (New Zealand); PAEC (Pakistan); MES and NSC (Poland); FCT (Portugal); MESTD (Serbia); MICIU/AEI and PCTI (Spain); MOSTR (Sri Lanka); Swiss Funding Agencies (Switzerland); MST (Taipei); MHESI and NSTDA (Thailand); TUBITAK and TENMAK (Turkey); NASU (Ukraine); STFC (United Kingdom); DOE and NSF (USA).

\hyphenation{Rachada-pisek} Individuals have received support from the Marie-Curie programme and the European Research Council and Horizon 2020 Grant, contract Nos.\ 675440, 724704, 752730, 758316, 765710, 824093, 101115353, 101002207, and COST Action CA16108 (European Union); the Leventis Foundation; the Alfred P.\ Sloan Foundation; the Alexander von Humboldt Foundation; the Science Committee, project no. 22rl-037 (Armenia); the Fonds pour la Formation \`a la Recherche dans l'Industrie et dans l'Agriculture (FRIA-Belgium); the Beijing Municipal Science \& Technology Commission, No. Z191100007219010 and Fundamental Research Funds for the Central Universities (China); the Ministry of Education, Youth and Sports (MEYS) of the Czech Republic; the Shota Rustaveli National Science Foundation, grant FR-22-985 (Georgia); the Deutsche Forschungsgemeinschaft (DFG), among others, under Germany's Excellence Strategy -- EXC 2121 ``Quantum Universe" -- 390833306, and under project number 400140256 - GRK2497; the Hellenic Foundation for Research and Innovation (HFRI), Project Number 2288 (Greece); the Hungarian Academy of Sciences, the New National Excellence Program - \'UNKP, the NKFIH research grants K 131991, K 133046, K 138136, K 143460, K 143477, K 146913, K 146914, K 147048, 2020-2.2.1-ED-2021-00181, TKP2021-NKTA-64, and 2021-4.1.2-NEMZ\_KI-2024-00036 (Hungary); the Council of Science and Industrial Research, India; ICSC -- National Research Centre for High Performance Computing, Big Data and Quantum Computing and FAIR -- Future Artificial Intelligence Research, funded by the NextGenerationEU program (Italy); the Latvian Council of Science; the Ministry of Education and Science, project no. 2022/WK/14, and the National Science Center, contracts Opus 2021/41/B/ST2/01369 and 2021/43/B/ST2/01552 (Poland); the Funda\c{c}\~ao para a Ci\^encia e a Tecnologia, grant CEECIND/01334/2018 (Portugal); the National Priorities Research Program by Qatar National Research Fund; MICIU/AEI/10.13039/501100011033, ERDF/EU, "European Union NextGenerationEU/PRTR", and Programa Severo Ochoa del Principado de Asturias (Spain); the Chulalongkorn Academic into Its 2nd Century Project Advancement Project, and the National Science, Research and Innovation Fund via the Program Management Unit for Human Resources \& Institutional Development, Research and Innovation, grant B39G670016 (Thailand); the Kavli Foundation; the Nvidia Corporation; the SuperMicro Corporation; the Welch Foundation, contract C-1845; and the Weston Havens Foundation (USA).
\end{acknowledgments}

\bibliography{auto_generated}
\cleardoublepage \appendix\section{The CMS Collaboration \label{app:collab}}\begin{sloppypar}\hyphenpenalty=5000\widowpenalty=500\clubpenalty=5000\input{HIG-23-014-public-authorlist.tex}\end{sloppypar}
\end{document}

%% file: HIG-23-014-public-authorlist.tex
\cmsinstitute{Yerevan Physics Institute, Yerevan, Armenia}
{\tolerance=6000
A.~Hayrapetyan, V.~Makarenko\cmsorcid{0000-0002-8406-8605}, A.~Tumasyan\cmsAuthorMark{1}\cmsorcid{0009-0000-0684-6742}
\par}
\cmsinstitute{Institut f\"{u}r Hochenergiephysik, Vienna, Austria}
{\tolerance=6000
W.~Adam\cmsorcid{0000-0001-9099-4341}, J.W.~Andrejkovic, L.~Benato\cmsorcid{0000-0001-5135-7489}, T.~Bergauer\cmsorcid{0000-0002-5786-0293}, K.~Damanakis\cmsorcid{0000-0001-5389-2872}, M.~Dragicevic\cmsorcid{0000-0003-1967-6783}, C.~Giordano, P.S.~Hussain\cmsorcid{0000-0002-4825-5278}, M.~Jeitler\cmsAuthorMark{2}\cmsorcid{0000-0002-5141-9560}, N.~Krammer\cmsorcid{0000-0002-0548-0985}, A.~Li\cmsorcid{0000-0002-4547-116X}, D.~Liko\cmsorcid{0000-0002-3380-473X}, I.~Mikulec\cmsorcid{0000-0003-0385-2746}, J.~Schieck\cmsAuthorMark{2}\cmsorcid{0000-0002-1058-8093}, R.~Sch\"{o}fbeck\cmsAuthorMark{2}\cmsorcid{0000-0002-2332-8784}, D.~Schwarz\cmsorcid{0000-0002-3821-7331}, M.~Shooshtari, M.~Sonawane\cmsorcid{0000-0003-0510-7010}, W.~Waltenberger\cmsorcid{0000-0002-6215-7228}, C.-E.~Wulz\cmsAuthorMark{2}\cmsorcid{0000-0001-9226-5812}
% ...
\par}
\vskip\cmsinstskip
\dag:~Deceased\\
$^{1}$Also at Yerevan State University, Yerevan, Armenia\\
$^{2}$Also at TU Wien, Vienna, Austria\\
% ...
